\documentclass[sigconf]{acmart}

\usepackage{amsmath}
\usepackage{rotating}
\usepackage{caption}
\usepackage{graphicx}
\usepackage{epstopdf}
\usepackage{epsfig,graphicx}
\usepackage{color}
\usepackage{url}
\usepackage{tabularx}
\usepackage{xfrac}
\usepackage{array}
\usepackage{subfigure}
\usepackage{multirow}
\usepackage{enumitem}
\usepackage{hyperref}
\usepackage{algorithm,algorithmic}
\usepackage{hhline}
 \usepackage{tikz}
\usepackage{booktabs}
\newcommand {\otoprule }{\midrule[\heavyrulewidth]}
\usepackage{adjustbox}
\usepackage{hyperref}
\definecolor{darkblue}{rgb}{0,0,.7}
\hypersetup{
     colorlinks   = true,
     citecolor    = blue,
     urlcolor     = blue,
     linkcolor    = blue
     }

\usepackage[section]{placeins}

\usepackage{sistyle}
\SIthousandsep{,}

\let\crnotice\mycrnotice%
\let\confname\myconfname%

\begin{document}
\clubpenalty=10000 
\widowpenalty = 10000
\newcommand{\tuan}[1]{{\textcolor{blue}{[\emph{\small Tuan: #1}]}}}
\newcommand{\khoi}[1]{{\textcolor{blue}{[\emph{\small Khoi: #1}]}}}

\newcommand\vpar{{\vspace*{1em}}}
\newcommand{\para}[1]{\noindent{\textbf{#1.}}}
\newcommand{\parai}[1]{\noindent{\textit{#1.}}}
\newcommand{\term}[1]{{\it {\small #1}}}

\newtheorem{mydef}{Definition}

\newcommand{\superscript}[1]{\ensuremath{^{\textrm{#1}}}}
\def\sharedaffiliation{\end{tabular}\newline\begin{tabular}{c}}
\def\ls{\superscript{1}}
\def\xls{\superscript{1,}}
\def\dfki{\superscript{2}}
\def\dk{\superscript{3}}




\title[On the Predictability of non-CGM Diabetes Data]{On the Predictability of non-CGM Diabetes Data for Personalized Recommendation}

\author{Tu Nguyen}
\affiliation{L3S Research Center}
\email{tunguyen@l3s.de}
\author{Markus Rokicki}
\affiliation{L3S Research Center}
\email{rokicki@l3s.de}
\begin{abstract}
With continuous glucose monitoring (CGM), data-driven models on blood glucose prediction have been shown to be effective in related work. However, such (CGM) systems are not always available, e.g., for a patient at home. In this work, we conduct a study on 9 patients and examine the \textit{online} predictability of data-driven (aka. machine learning) based models on patient-level blood glucose prediction; with measurements are taken only periodically (i.e., after several hours). To this end, we propose several \textit{heuristic}-based and \textit{statistical}-based post-prediction methods to account for the noise nature of these data, that yields marginally significant improvements to the performance of the overall system.
\end{abstract}
\maketitle
\section{Introduction}
Diabetes mellitus has been a major and global problem for a long time, as it is report that there are over 400 million patients over the world~\footnote{\url{https://www.diabetes.co.uk/diabetes-prevalence.html}}. The knowledge of glucose concentration in blood is a key aspect in the diagnosis and treatment of diabetes. The use of signal
processing techniques on glucose data started a long time ago, when glucose time-series in a given
individual could be obtained in lab study from samples drawn in the blood at a sufficiently high rate.  In particular, related work employed not only linear (e.g.,
correlation and spectrum analysis, peak detection), but also nonlinear (e.g., approximate entropy)
methods to investigate oscillations present in glucose (and insulin) time-series obtained, during
hospital monitoring, by drawing blood samples every 10-15 min for up to 48 h~\cite{sparacino2010smart}. In these settings, long term (e.g., days or months) studies resorted to self-monitoring blood glucose (SMBG) data, i.e., approx. 3 samples per day obtained by the patient herself by using fingerstick glucose meters. The retrospective analysis of SMBG time-series was used by physicians, together with the information taken from the `patient's diary` (e.g., insulin dosage, meals intake, physical exercise) and some glycaemic indexes (typically HbA1c), to assess glucose control and the effectiveness of a particular therapy~\cite{sparacino2010smart}. 

With the support of continuous glucose monitoring (CGM) sensors, the development of new strategies for the treatment of diabetes has been accelerated in recent years. In particular, CGM sensors can be injected into `online` recommender systems that are able to
generate alerts when glucose concentration is predicted to exceed the normal range thresholds. Recently, there has been a lot of complex data-driven prediction models~\cite{eren2009estimation,plis2014machine,contreras2017personalized,fiorini2017data} that are built based on the CGM data, and have been shown to be effective. These data-driven models, or machine learning/deep learning are data-hungry, hence, its performance on \textbf{sparse / non-continuous data} is still a question. CGM data are still not always available for all diabetic patients for many reasons~\footnote{\url{http://time.com/4703099/continuous-glucose-monitor-blood-sugar-diabetes/}}; while a personalized or \textit{patient-level} model that are trained on the same patient's data is essential. In this work, we examine the performance of these machine leaning approaches on our real, limited data of a group of diabetic patients. Our contributions are two-fold: (1) we provide a quantitative study on the predictability of machine learned models on limited and sparse data; (2) we propose a prediction system that is robust on noisy data (based on prediction interval).
\section{Dataset Overview}
The data collection study was conducted from end of February to beginning of April 2017  and includes 9 patients who were given specially prepared smartphones. Measurements on carbohydrate consumption, blood glucose levels, and insulin intake were made with the Emperras Esysta system~\footnote{\url{https://www.emperra.com/en/esysta-product-system/}}. Measurements on physical activities were obtained using the Google Fit app. We use only steps information (number of steps) for our study.

We describe briefly here some basic patient information. Half of the patients are female and ages range from 17 to 66, with a mean age of 41.8 years. Body weight, according to BMI (Body mass index), is normal for half of the patients, four are overweight and one is obese. The mean BMI value is 26.9. Only one of the patients suffers from diabetes type 2 and all are in ICT therapy~\footnote{describes as a model of an insulin therapy for the diabetics with two different types of insulin.}. In terms of time since being diagnosed with diabetes, patients vary from inexperienced (2 years) to very experienced (35 years), with a mean value of 13.9 years. We anonymize the patients and identify them by IDs (from 8 to 17, we do not have information for patient 9).

\subsection*{Frequency of Measurements}
We give an overview of the number of different measurements that are available for each patient. The study duration varies among the patients, ranging from 18 days, for patient 8, to 33 days, for patient 14. 
Likewise, the daily number of measurements taken for carbohydrate intake, blood glucose level and insulin units vary across the patients. 
The median number of carbohydrate log entries vary between 2 per day for patient 10 and 5 per day for patient 14. 
Median number of blood glucose measurements per day varies between 2 and 7. 
Similarly, insulin is used on average between 3 and 6 times per day. 
In terms of physical activity, we measure the 10 minute intervals with at least 10 steps tracked by the google fit app.
This very low threshold for now serves to measure very basic movements and to check for validity of the data.
Patients 11 and 14 are the most active, both having a median of more than 50 active intervals per day (corresponding to more than 8 hours of activity).
Patient 10 on the other hand has a surprisingly low median of 0 active 10 minutes intervals per day, indicating missing values due to, for instance, not carrying the smartphone at all times. 

\subsection*{Measurements per Hour of Day}
Figure~\ref{fig:hists_p13_p14} show measurements of blood glucose, carbohydrates and insulin per hour of day for patient 13 and 14. 
Overall, the distribution of all three kinds of values throughout the day roughly correspond to each other.
In particular, for most patients the number of glucose measurements roughly matches or exceeds the number of rapid insulin applications throughout the days.
Notable exceptions are patients 14, 15, and 17 (figures excluded).
For patient 14, in the evening the number of meals and rapid insulin applications match but exceed the number of blood glucose measurements by far. 
Patient 17 has more rapid insulin applications than glucose measurements in the morning and particularly in the late evening. 
For patient 15, rapid insulin again slightly exceeds the number of glucose measurements in the morning. Curiously, the number of glucose measurements match the number carbohydrate entries -- it is possible the discrepancy is a result of missing (glucose and carbohydrate) measurements. We further show the blood glucose distribution of each patient in Figure~\ref{fig:bgdis}. The different lengths of the interquartile range for each distribution also reflects the difficulty of prediction problem on different patients.

\begin{figure}[h]
\includegraphics[width=0.7\columnwidth]{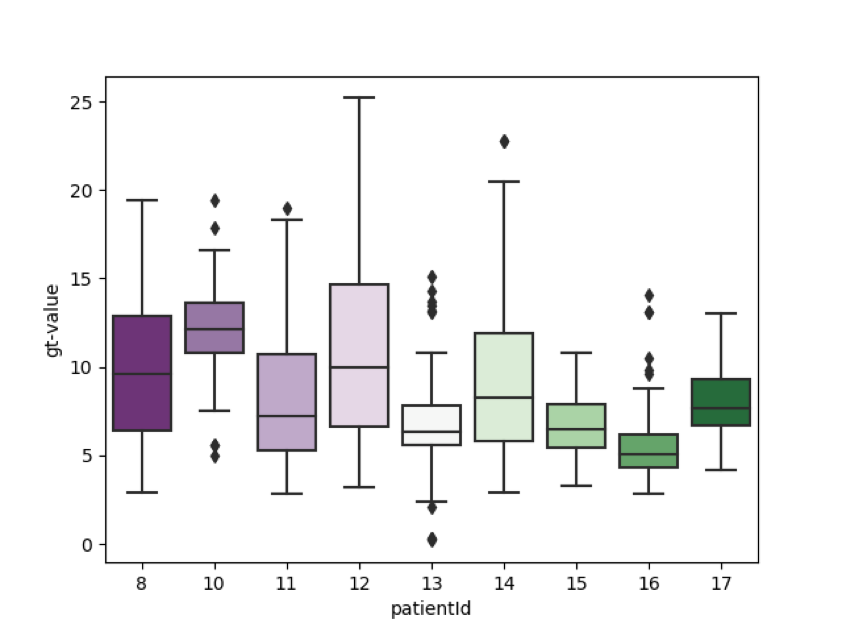}
\caption{Blood glucose distribution for each patient.}
\label{fig:bgdis}
\end{figure}

\begin{figure}[h]
\includegraphics[width=0.9\columnwidth]{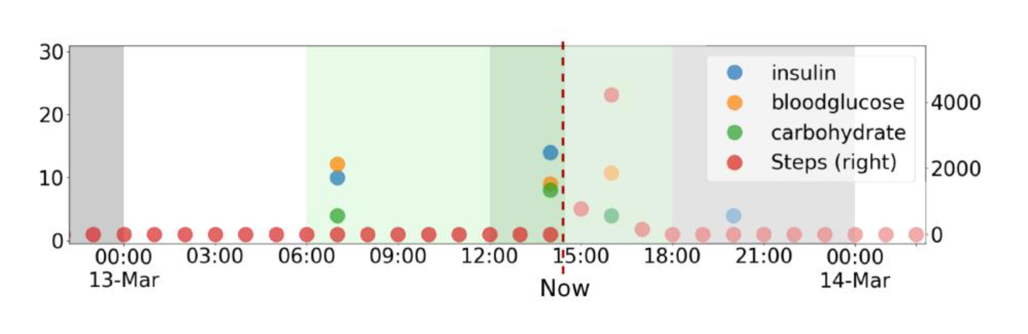}
\caption{Blood glucose prediction scenario.}
\label{fig:scenario}
\end{figure}
\begin{figure}[h]
\subfigure[P13 Glucose]{
\centering
\includegraphics[width=0.45\columnwidth]{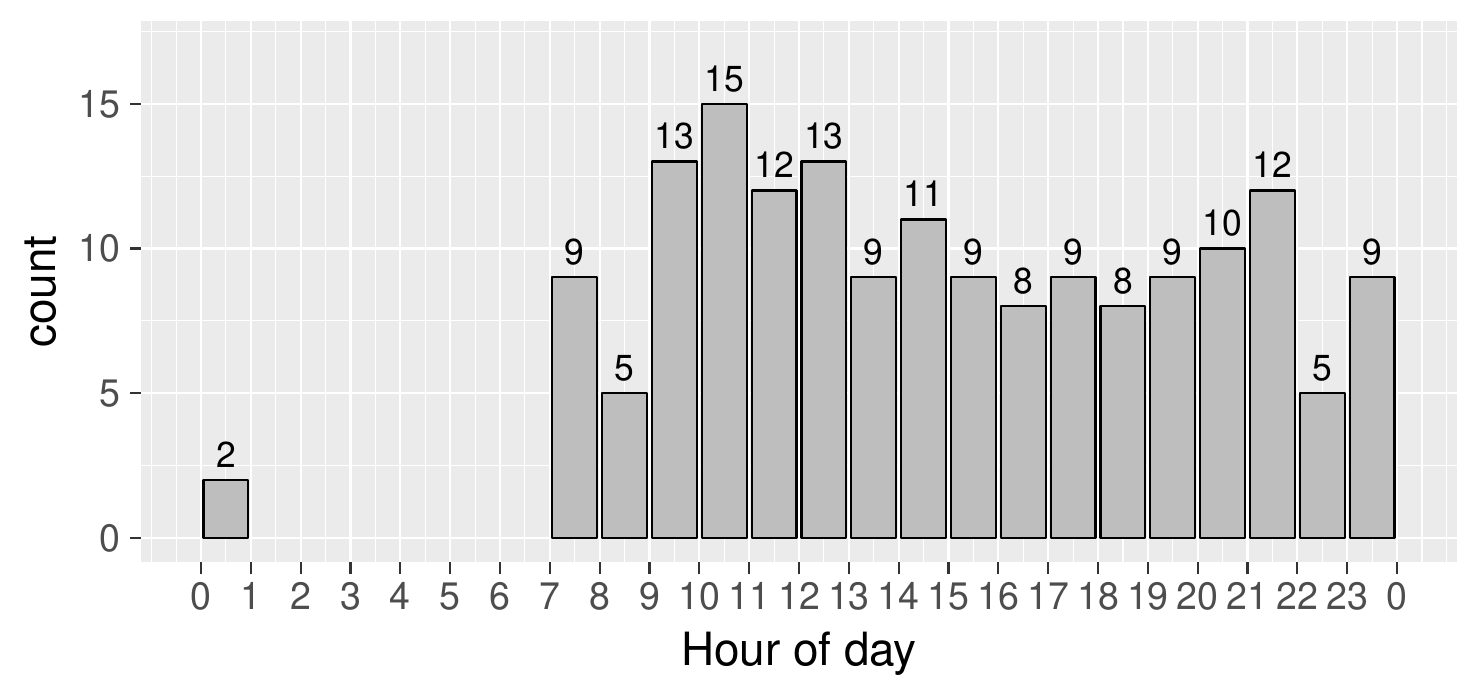}
}
\subfigure[P14 Glucose]{
\centering
\includegraphics[width=0.45\columnwidth]{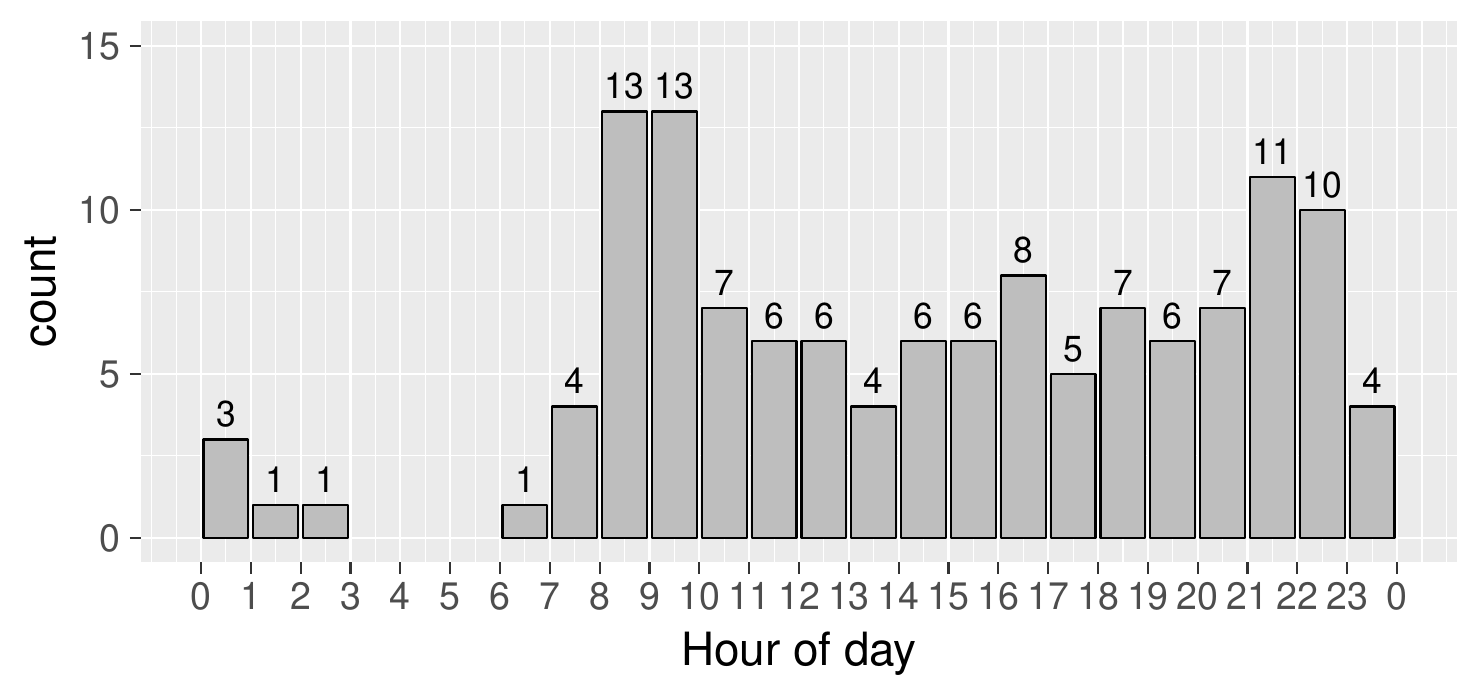}
}\\
\subfigure[P13 Carbohydrates]{
\centering
\includegraphics[width=0.45\columnwidth]{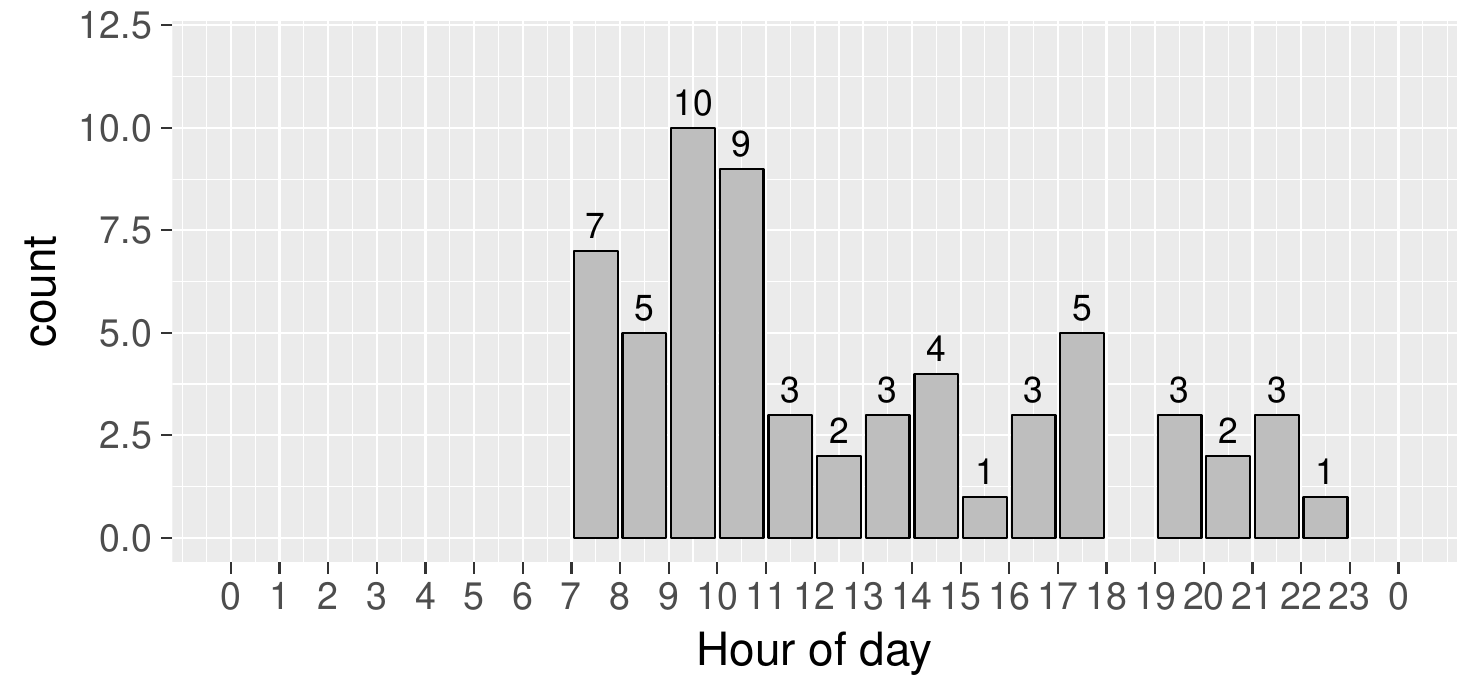}
}
\subfigure[P14 Carbohydrates]{
\centering
\includegraphics[width=0.45\columnwidth]{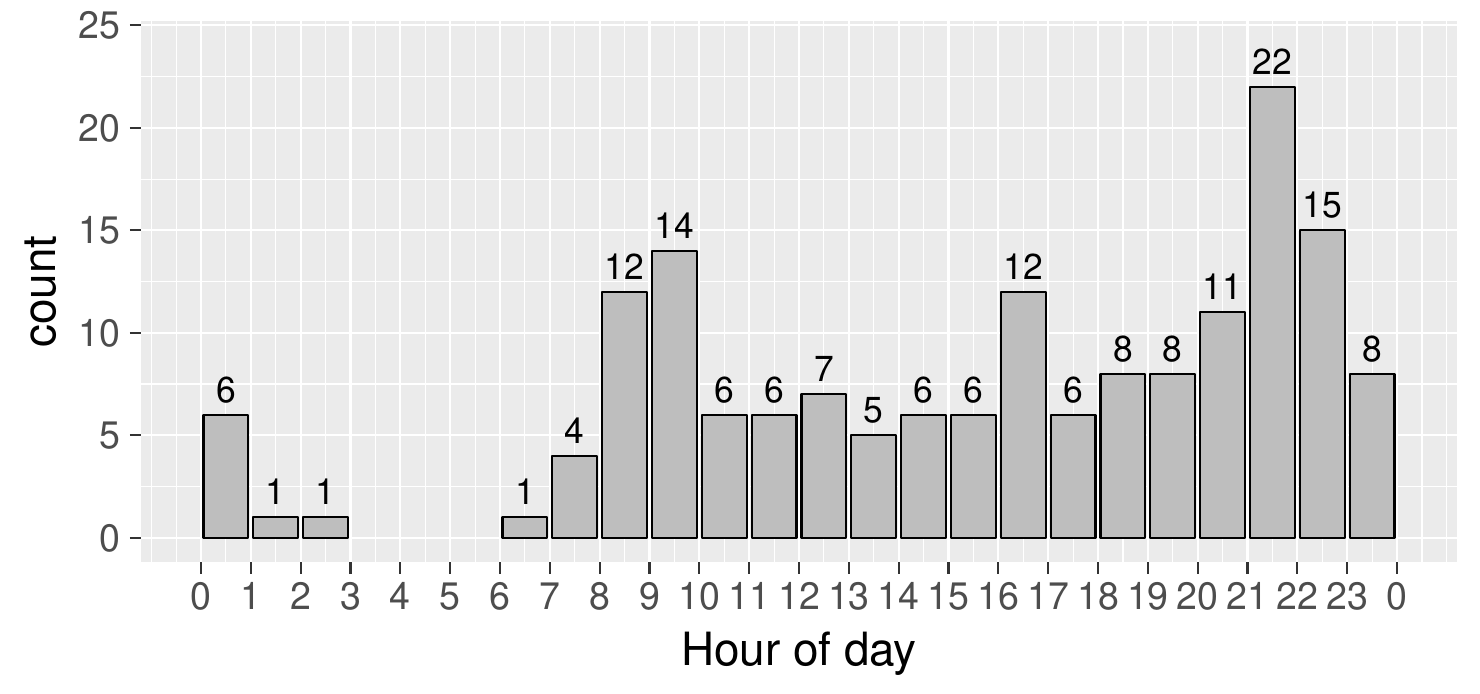}
}\\
\subfigure[P13 Insulin]{
\centering
\includegraphics[width=0.45\columnwidth]{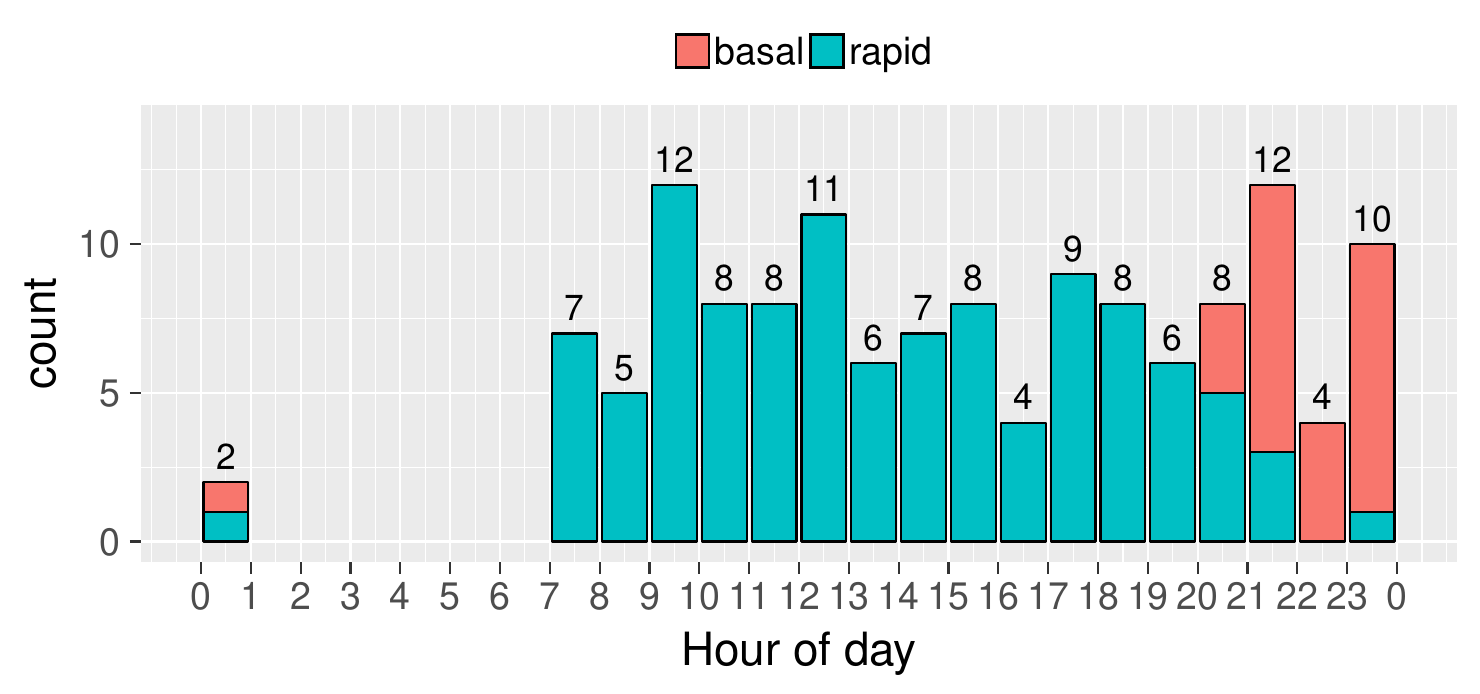}
}
\subfigure[P14 Insulin]{
\centering
\includegraphics[width=0.45\columnwidth]{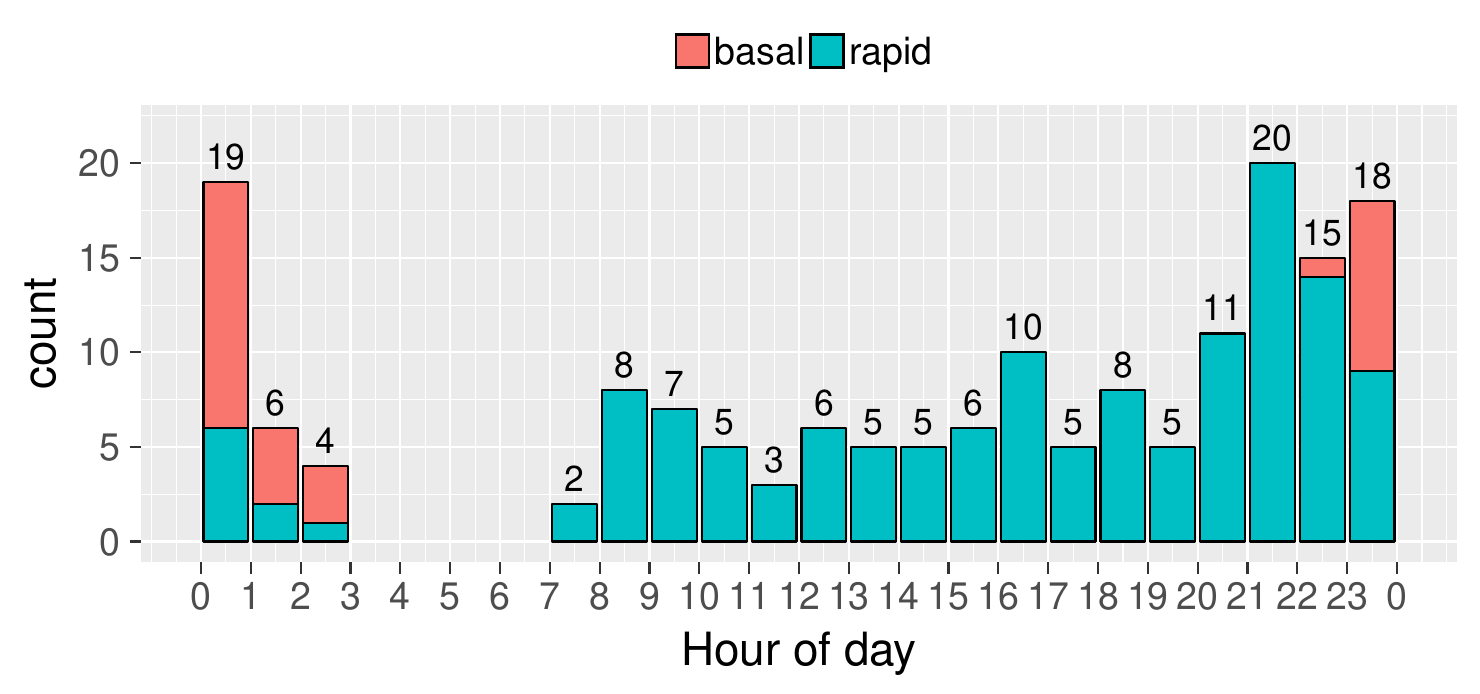}
}
\caption{Glucose, carbohydrate and insulin values per hour of day for patients 13 and 14.}
\label{fig:hists_p13_p14}
\end{figure}

\section{Prediction}
Our first approach to blood glucose prediction is based on a regression type form of time series prediction. 
Given historical blood glucose data, we learn a model that predicts future glucose values based on a representation of the current situation (including the recent past), 
using information on patient context, recent insulin applications, carbohydrate intake, and physical activity levels.

\subsection{Setup}
\paragraph*{Prediction task}
Our prediction task is a time series prediction of blood glucose values (in mmol/L) with a prediction horizon of 1 hour.
Consequently, we can construct a data instance for each glucose measurement found in the dataset and use all information available up until 1 hour before the measurement for predicting the glucose value (c.f., Figure~\ref{fig:scenario}).

\paragraph*{Evaluation Protocol}
Performance is evaluated on a per patient basis. In addition, we average performance over patients to get an overview.
For each patient, we consider the first 66\% of blood glucose measurements as training data to learn the models and the last 34\% as test data to evaluate prediction performance.

\paragraph*{Performance Measures}
Prediction performance is measured in terms of median absolute error (MdAE), root mean squared error (RMSE) and symmetric mean absolute percentage error (SMAPE).
Given are ground truth values $y_i$ and predictions $\hat{y}_i$, with $i \in [1,n]$. 
Median absolute error measures the median error made and is defined as $$
\operatorname{MdAE} = \underset{i}{\operatorname{median}}(|\hat{y}_i - y_i| ).
$$

Root mean squared error weighs larger errors more heavily and is defined as
$$
\operatorname{RMSE} = \sqrt{\frac{\sum^n_{i=1} (\hat{y}_i - y_i)^2 }{n}}.
$$

Symmetric mean absolute percentage error relates prediction errors to predicted values. It is defined as
$$
\operatorname{SMAPE} = \frac{100\%}{n} \sum^n_{i=1}\frac{|\hat{y}_i - y_i|}{(|y_i| + |\hat{y}_i|)/2}.
$$
Note that this gives a result between 0\% and 200\%. Further, the measure penalizes a) deviating for low values and b) over-forecasting.

\subsection{Algorithms}
\paragraph*{Simple Baselines}
As standard simple baselines, we use the last value observed one hour before the value that is being predicted (\emph{Last}) and the arithmetic mean of glucose values in the training set.

\paragraph*{Context-AVG} 
As a more advanced baseline, we use a (temporal) context weighted average of previous glucose values.  
As our analysis showed differences in glucose values according to time of the day, we weigh previous glucose values base on temporal proximity, weighted exponentially decreasing in the difference of time of day.
\paragraph*{Long-short-term-memory}. LSTM is a recurrent neural network model that effectively accounts for the long-term sequence dependence among glucose inputs.

\paragraph*{RandomForest}
The Random Forest Regressor (RF) is a meta estimator that learns an ensemble of regression trees~\cite{breiman2001random}, averaging the output of individual regression trees to perform the prediction. We use a standard value of 500 estimators, as well as a minimal leaf size of 4 for the individual trees to reduce overfitting of the individual models.

\paragraph*{ExtraTrees}
The Extra-Trees Regressor (ET) is a variation on RandomForest that uses a different base learner: Extremely randomized trees~\cite{geurts2006extremely}. In contrast to regular regression trees, best split values per feature are chosen randomly. We use 300 estimators and a minimum leaf size of 2.

\subsection{Overall Results}
In this section we report aggregate results, averaged over all patients. Table~\ref{tab:reg_all} shows regression performance averaged over all patients.
Performance is based on 42 test instances on average. The simple baselines \emph{Last} and \emph{AVG} achieve median errors of 3.3 and 2.5 mmol/L.
Weighing previous glucose values based on time of the day (\emph{Context-AVG}) improves average median errors to 2.28 mmol/L.
The Extra-Trees Regressor achieves the lowest MdAE of 2.16 and similarly slightly outperforms Context-AVG in terms of RMSE and SMAPE.
In comparison to predicting the arithmetic mean (\emph{AVG}), however, RMSE does not improve by much (12.15 vs 12.96), indicating that the ensemble is not able to predict extreme errors well on average.
We additionally report the performance of a neural-network based model, the Long-short-term-memory (LSTM), trained with 10 and 100 epochs. LSTM seems to be quite stable for MdAE but varies substantially for RMSE and SMAPE. The performance of LSTM actually gets much worse after 100 epochs, that indicates the prone to \textit{overfitting}. This show the instability of the model towards our dataset, and thus we do not consider the LSTM results for model comparison in Table~\ref{tab:reg_all}. 

\begin{table}[h]
  \centering
  \small
  \begin{tabular}{lrrr}
    \otoprule
    Method & MdAE & RMSE & SMAPE \\
    \otoprule
    Last & 3.28 & 25.71 & 40.96 \\
    AVG & 2.51 & 12.96 & 31.42 \\
    Context-AVG & 2.28 & 12.53 & 29.71 \\
    ARIMA & 2.40 & 13.88 & 31.61 \\
    LSTM (10 iter) & \textit{ 2.24} & \textit{10.41} & \textit{29.02} \\
    LSTM (100 iter) & \textit{2.76} & \textit{19.24} & \textit{35.64} \\ 
    RandomForest & 2.27 & \textbf{12.05} & 29.98 \\
    Extremely (randomized) Trees & \textbf{2.16} & 12.15 & \textbf{29.56} \\
    \bottomrule
  \end{tabular}
\vspace{.25em}
  \caption{Overall regression performance averaged over all patients. Best performance per measure is marked in bold (results in italic are not considered for comparison).}
  \label{tab:reg_all}
\end{table}

\section{Prediction confidence}
In this section, we study the \textit{prediction confidence} of our best performed prediction tree-based bagging models, RandomForest and ExtraTrees. This would, to an extent, facilitate us to answer an important question, \textit{\textbf{when} the bagged model can omit reliable predictions?} Typically, the prediction confidence can be associated to the \textit{sampling variance} introduced by the bagging process.

\paragraph{Prediction intervals} 
When looking at two regression models, while the model predictions could be similar, confidence in them would vary if we look at the training data, a less and more spread out data could bring a low confidence. Hence, a prediction returning a single value (typically meant to minimize the squared error) likewise does not relay any information about the underlying distribution of the data or the range of response values. We hence, leverage the notion of \textbf{prediction confidence} to supplement for the noisy data and enhance the end model, in the sense that the model has the capability to \textit{abstain} from making predictions at certain times when its prediction \textit{confidence} is low.

A \textit{prediction interval} or \textit{confidence interval} is an estimate of an interval into which the future observations will fall with a given probability. In other words, it can quantify our confidence or certainty in the prediction. Unlike confidence intervals from classical statistics, which are about a parameter of population (such as the mean), prediction intervals are about individual predictions~\cite{datadive}.

Formally, follow~\cite{wager2014confidence}, suppose that we have training examples $Z_1 = \left(x_1, y_1\right), \ldots, Z_n = \left(x_n, y_n\right)$, an input $x$ to a prediction problem, and a base decision tree learner $\hat{\theta}\left(x\right) = t\left(x; Z_1, \ldots, Z_n\right)$.

With bagging, we aim to stabilize the base learner $t$ by re-sampling the training data. In our case, the bagged version of $\hat{\theta}\left(x\right)$ is defined as
\begin{equation}
    \hat{\theta}_\infty\left(x\right) = \mathbb{E}^*\left[t(x; Z^*_1, \ldots, Z^*_n)\right], 
\label{eq:bagged_estimator}
\end{equation}
where the $Z^*_i$ are drawn independently with replacement from the original data (i.e., they form a bootstrap sample). The expectation $\mathbb{E}^*$ is taken with respect to the bootstrap measure.

The expectation in \eqref{eq:bagged_estimator} can be estimated by Monte Carlo as:
\begin{equation}
\hat{\theta}_B\left(x\right) = \frac{1}{B} \sum_{b=1}^{B} t^*_b\left(x\right), 
\label{eq:bagged_estimator_mc}
\end{equation}
where $t^*_b\left(x\right) = t\left(x; Z^*_{b1}, \ldots, Z^*_{bn}\right)$ and the $Z^*_{bi}$ are elements in the $b$-th bootstrap sample. As $B \rightarrow \infty$, we recover the perfectly bagged estimator $\hat{\theta}_\infty\left(x\right)$.

Subsequently, the sampling variance of the bagged learners is:

\begin{equation}
v\left(x\right) = \text{var}\left[\hat{\theta}_{\infty}\left(x\right)\right],
\end{equation}

The accuracy of variance estimation is frequently influenced by the inherent Monte Carlo noise that emerges due to the limited number of bootstrap learners. In this study, we utilize a bias-corrected technique, as introduced in~\cite{wager2014confidence}. Subsequently, we present the variance estimation in two modes: (1) \textit{bias} and (2) \textit{bias correction}.


\subsection{Regression evaluation}

We report here the prediction variability evaluation across all patients for the regression task. Figure~\ref{fig:ebr} show the error bars using unbiased variance for all patients. We then show in Figures~\ref{fig:tns} the error bar graphs for patient 8 in an \textit{incremental} training size setting -- meaning that we keep the same actual test set, but training on only part of the training data. E.g., $1/4$ training data indicates that we `look back' on only $1/4$ of the available past data. The more dots that near the diagonal show the more `accurate' is our prediction model. And the error bars show the `confidence' interval. Figure~\ref{fig:iebr8a} indicates the high `confidence' in the predictions with little training data, yet the dots are far away from the diagonal.

\begin{figure}[ht]
\subfigure[Patient 8]{
\centering
\includegraphics[width=0.3\columnwidth]{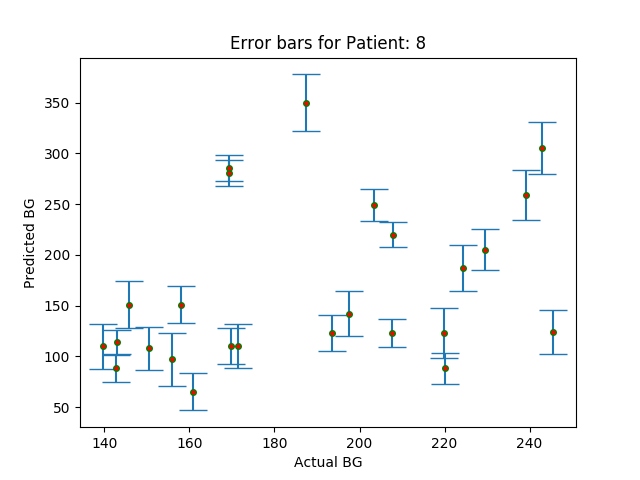}
}
\subfigure[Patient 10]{
\centering
\includegraphics[width=0.3\columnwidth]{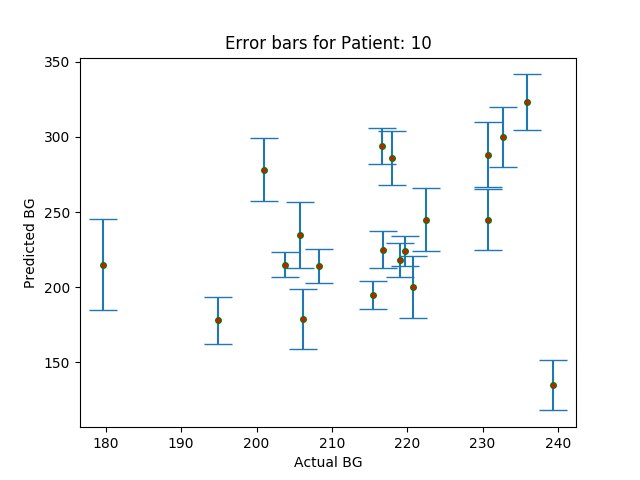}
}
\subfigure[Patient 11]{
\centering
\includegraphics[width=0.3\columnwidth]{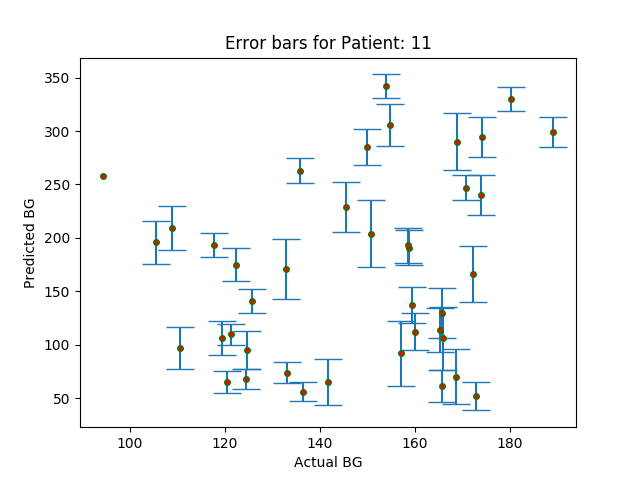}
}\\
\subfigure[Patient 12]{
\centering
\includegraphics[width=0.3\columnwidth]{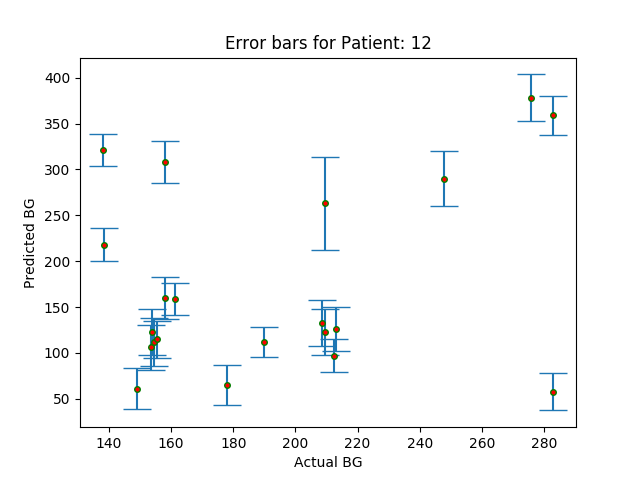}
}
\subfigure[Patient 13]{
\centering
\includegraphics[width=0.3\columnwidth]{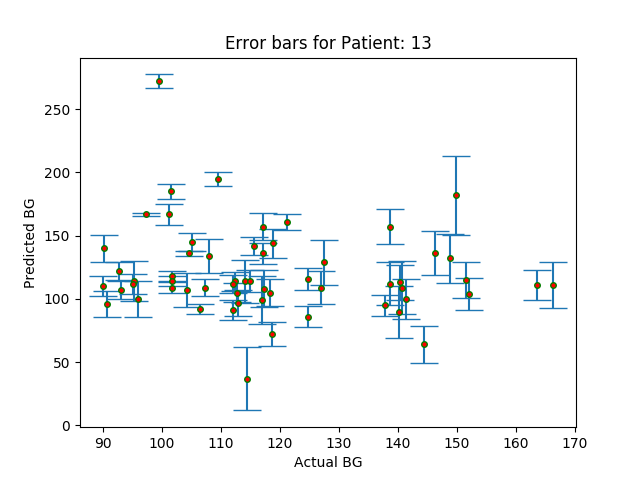}
}
\subfigure[Patient 14]{
\centering
\includegraphics[width=0.3\columnwidth]{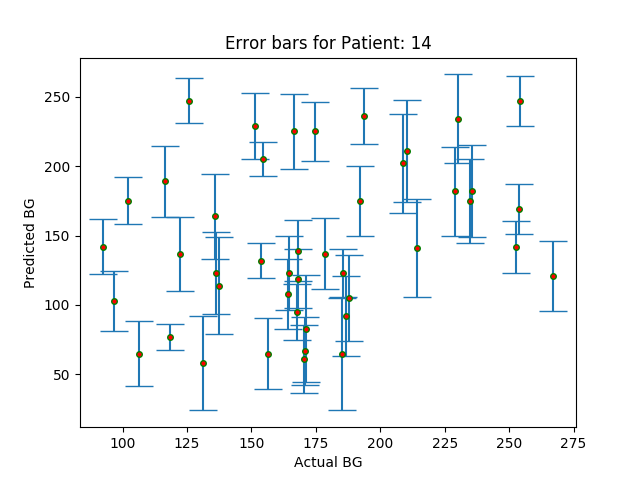}
}\\
\subfigure[Patient 15]{
\centering
\includegraphics[width=0.3\columnwidth]{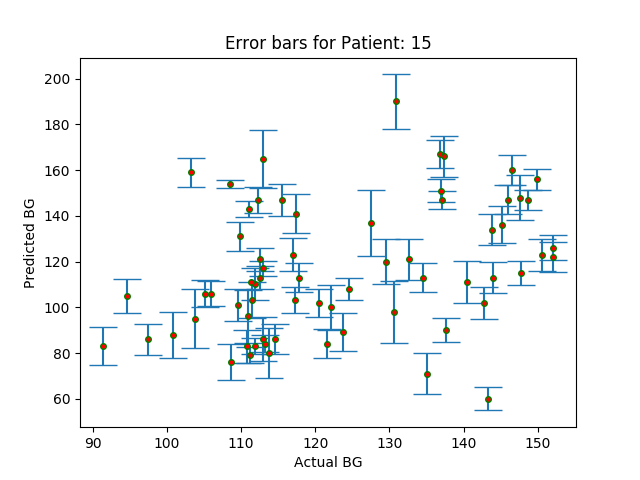}
}
\subfigure[Patient 16]{
\centering
\includegraphics[width=0.3\columnwidth]{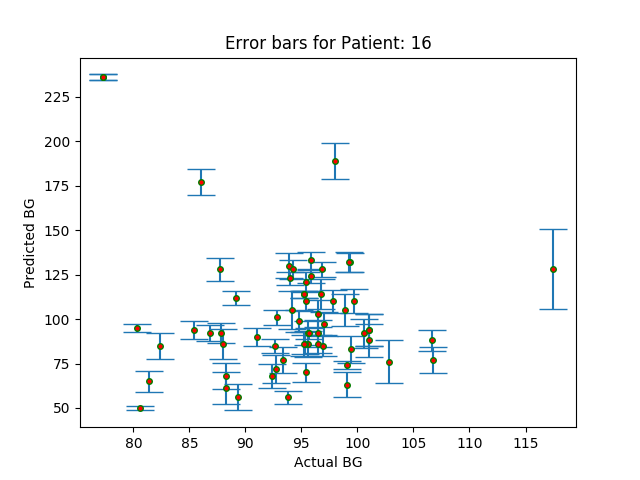}
}
\subfigure[Patient 17]{
\centering
\includegraphics[width=0.3\columnwidth]{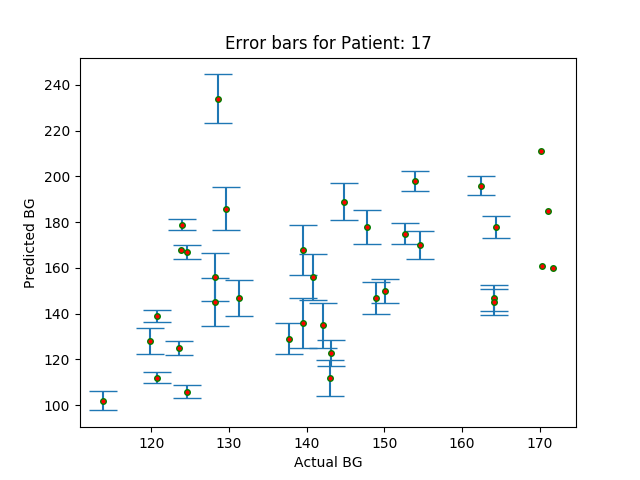}
}
\caption{Error bar graphs for predicted BG using unbiased variance.}
\label{fig:ebr}
\end{figure}

\begin{figure}[ht]
\subfigure[1/4 training data]{
\centering
\includegraphics[width=0.4\columnwidth]{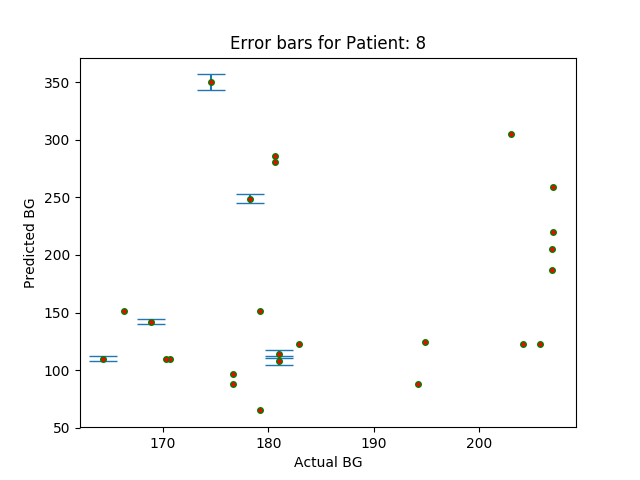}
\label{fig:iebr8a}
}
\subfigure[2/4 training data]{
\centering
\includegraphics[width=0.4\columnwidth]{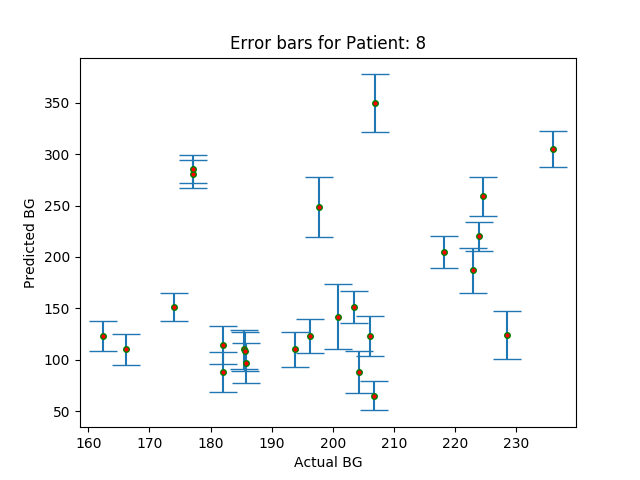}
}\\
\subfigure[3/4 training data]{
\centering
\includegraphics[width=0.4\columnwidth]{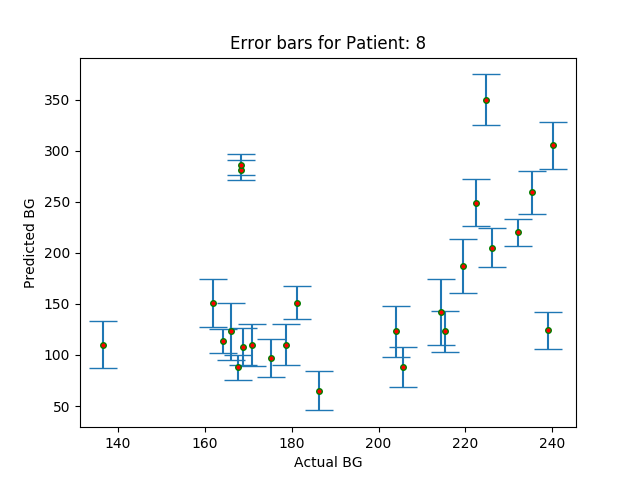}
}
\subfigure[4/4 training data]{
\centering
\includegraphics[width=0.4\columnwidth]{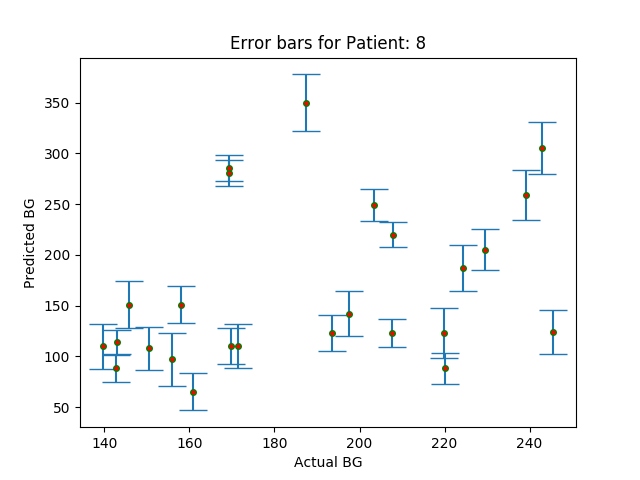}
}
\caption{Incremental training size - error bar graphs for predicted BG using unbiased variance for patient 8.}
\label{fig:tns}
\end{figure}

\subsubsection{when to predict: on the training size evaluation.}
To answer this question, we set up an evaluation setting with increasing size of number of instances, order by time. Each training point is evaluated by leave-one-out validation.  We show in Figure~\ref{fig:lo1} the results for patient 8. The general conclusion is the that the more training data, the better the performance is, as witness for patient 13, 15 or 17. However, the results for such patients e.,g patient 8, 11 or 16 show that the training size increment could also bring more noise and decrease the results. We envision that it could because the learned model is not stabilized yet with the limited number of instances in our experiment. In addition, training size is not the only factor to decide when to predict. We hence move on to examine the other two factors: (1) model stability - via std. dev. and (2) prediction confidence toward coming instances.

\subsubsection{when to predict: on the model stability.}
To answer this question, we measure the stability of the model by the standard deviation of the k-fold cross validation with incremental training size. Figure~\ref{fig:std} indicate on MAE and RMSE metrics, the model seems to be more stabilized with the more number of training data. This is a good indicator for the \textit{when to predict} questions.

\subsubsection{when to predict: on the prediction confidence.}
We show in Figure~\ref{fig:disconfbias} and Figure~\ref{fig:disconf} the confidence distribution at each run of the 5-fold CV for different patients based on \textit{bias} and \textit{no-bias} confidences respectively. The results show the confidence distributions are rather similar across different run, indicating that the temporal order of the instances does not impact much on the model performance. Base on the distribution, we move on the the threshold parameter tuning for the data filtering using confidence interval. The idea is to answer the question, "if we filter low confidence instances (high confident interval), will the model perform better?"

The answer somehow is depicted in Figure~\ref{fig:cthresh}. For some patients i.e., patient 10, 13, the filtering technique substantially enhance the model performances on MAE and RMSE (not shown) metrics. It is witness that the \textit{biased} confidence measure somewhat works better than \textit{non-biased} one across patients. However, for some patients i.e., patient 8 it seems does not bring any effects. 

We move on to experiment with \textbf{filtering} instances that we empirically witness that it seems lacking of preditable context within the training data. They are the BG measurements at \textbf{night}. We then attempt to filter those out for prediction. Even though slightly improving for some (c.f., Figure~\ref{fig:night}), overall the filtering attempt does not make significant difference, indicating that our model learns it better than we expect.

\subsubsection{when to predict: combined factors.} Figure~\ref{fig:fcomb} show some highlighted combined filtering techniques. In general, combining the aforementioned factors together does improve the model performance. However, the combination is not straightforward, e.g., \textit{confidence interval} filtering lower the performance at the starting time when the model is unstable aka. \textit{cold start}. Hence, there is not enough evidence for us to make a \textit{hard} decision. The more \textit{trial-and-error} attempts on the fly or a bigger dataset however will be at ease to be built on these as a foundation.

\subsection{Overall results with Filtering methods} We show in Table~\ref{tab:filtersum} the overall results of our models with different filtering approaches for all patients. We use 2 different filtering approaches: (1) Sanity filter, heuristics (e.g., remove out wrongly input measurement or moments when the last glucose level input is too far) that remove noise
and (2) Stability filter: prediction confidence (std. dev is not needed when the training size is large enough). The results show that the \textit{stability filter} (based on bias and bias-corrected) achieve the best performance, without the need of human efforts on sanity filter. Sole stability filter also provide more predictions (avg. 24) than other filtering combination.

\begin{figure}[ht]
\subfigure[Patient 8]{
\centering
\includegraphics[width=0.3\columnwidth]{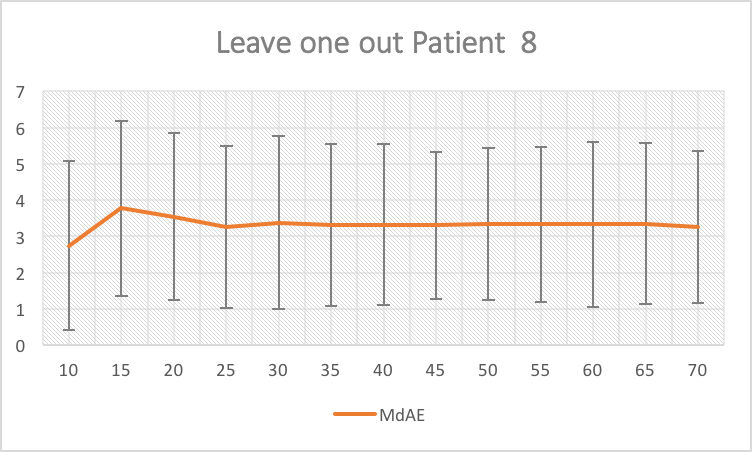}
}
\subfigure[Patient 10]{
\centering
\includegraphics[width=0.3\columnwidth]{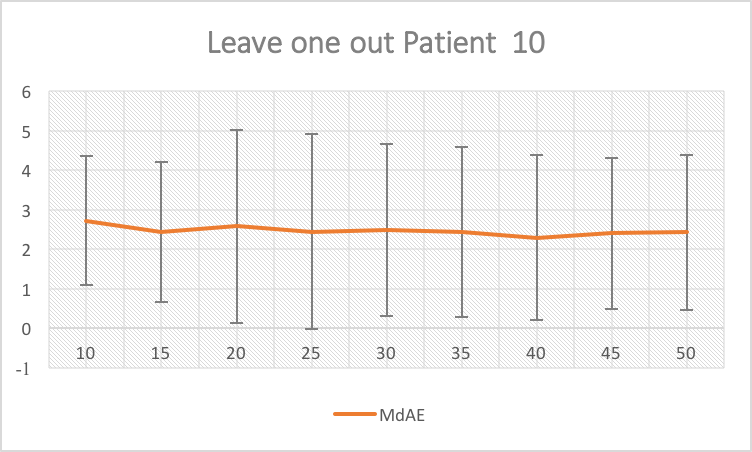}
}
\subfigure[Patient 11]{
\centering
\includegraphics[width=0.3\columnwidth]{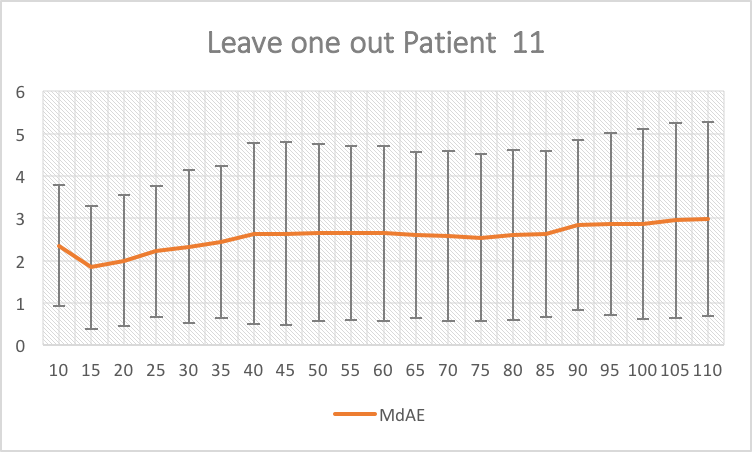}
}\\
\subfigure[Patient 12]{
\centering
\includegraphics[width=0.3\columnwidth]{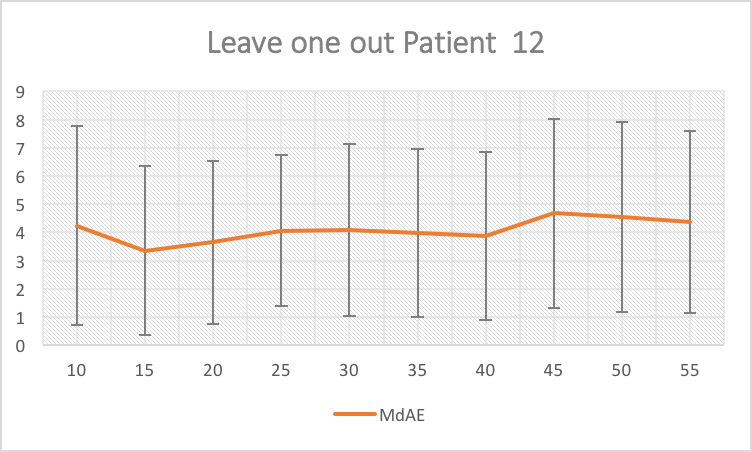}
}
\subfigure[Patient 13]{
\centering
\includegraphics[width=0.3\columnwidth]{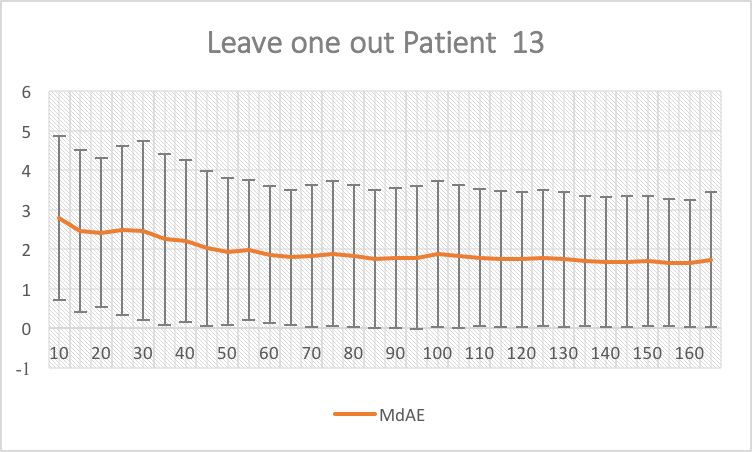}
}
\subfigure[Patient 14]{
\centering
\includegraphics[width=0.3\columnwidth]{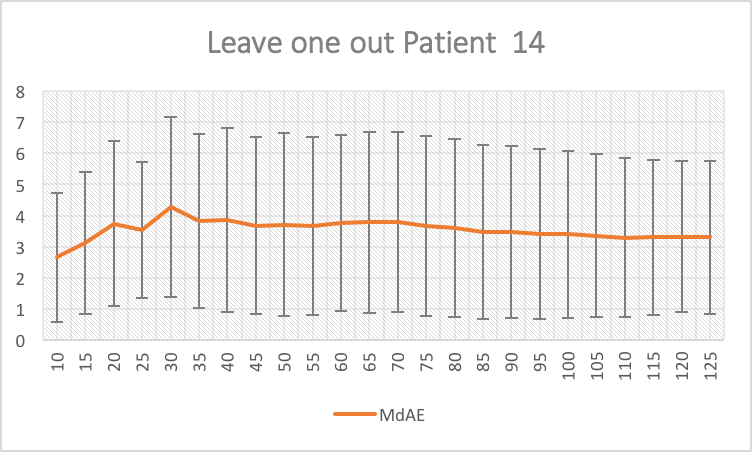}
}\\
\subfigure[Patient 15]{
\centering
\includegraphics[width=0.3\columnwidth]{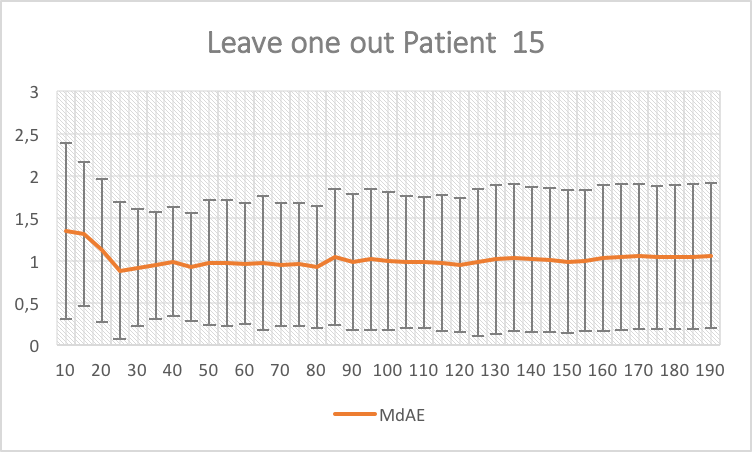}
}
\subfigure[Patient 16]{
\centering
\includegraphics[width=0.3\columnwidth]{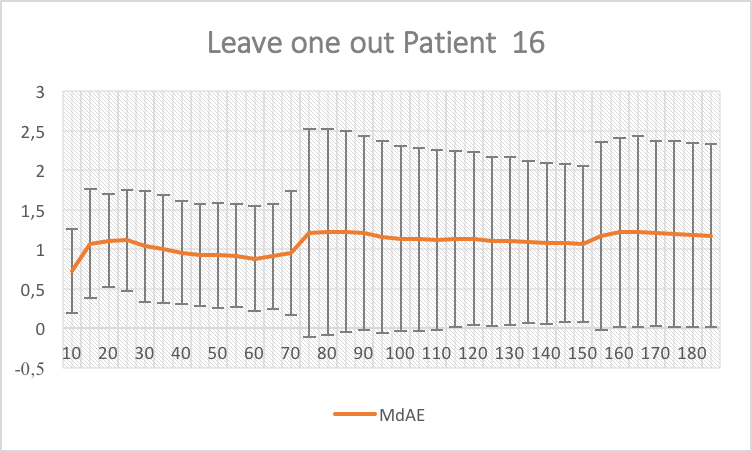}
}
\subfigure[Patient 17]{
\centering
\includegraphics[width=0.3\columnwidth]{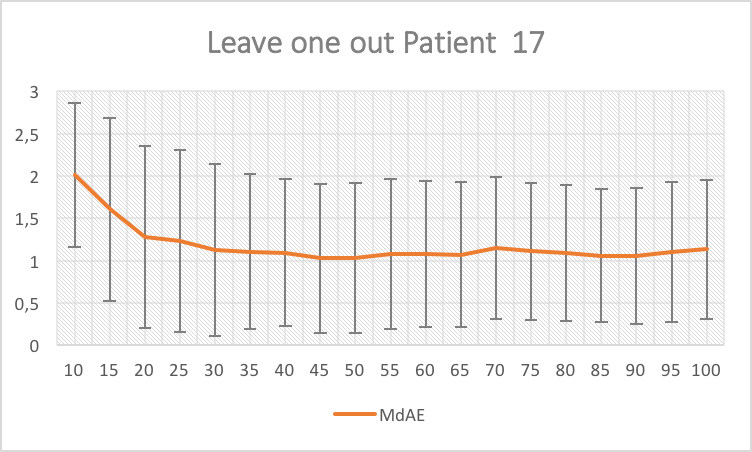}
}
\caption{Leave-one-out cross validation with incremental training size.}
\label{fig:lo1}
\end{figure}

\begin{figure}[ht]
\subfigure[Patient 8]{
\centering
\includegraphics[width=0.45\columnwidth]{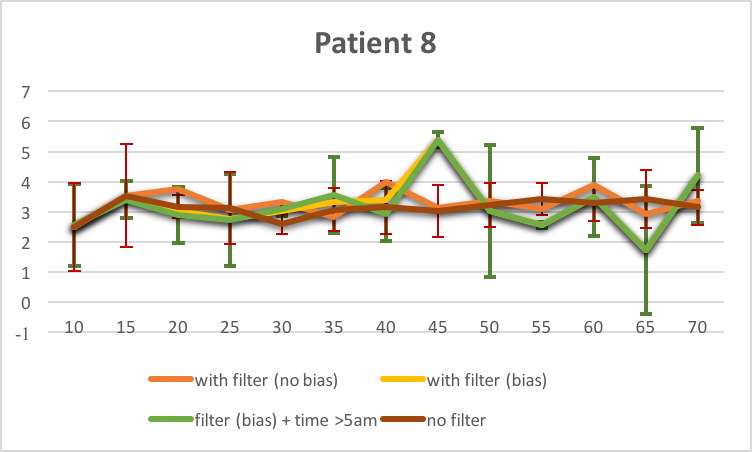}
}
\subfigure[Patient 13]{
\centering
\includegraphics[width=0.45\columnwidth]{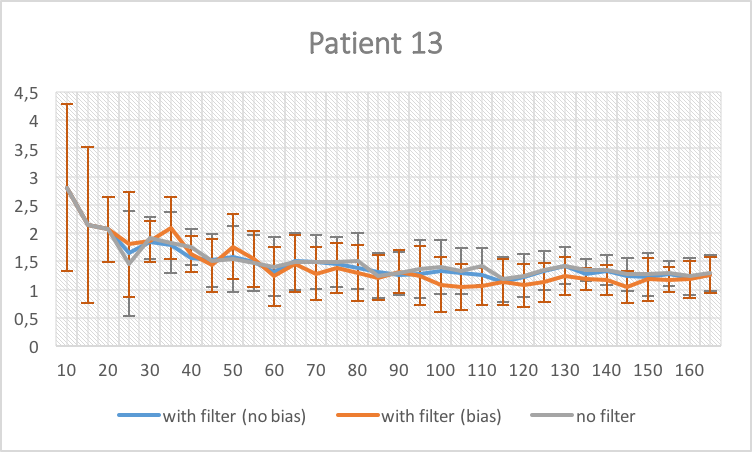}
}\\
\subfigure[Patient 15]{
\centering
\includegraphics[width=0.45\columnwidth]{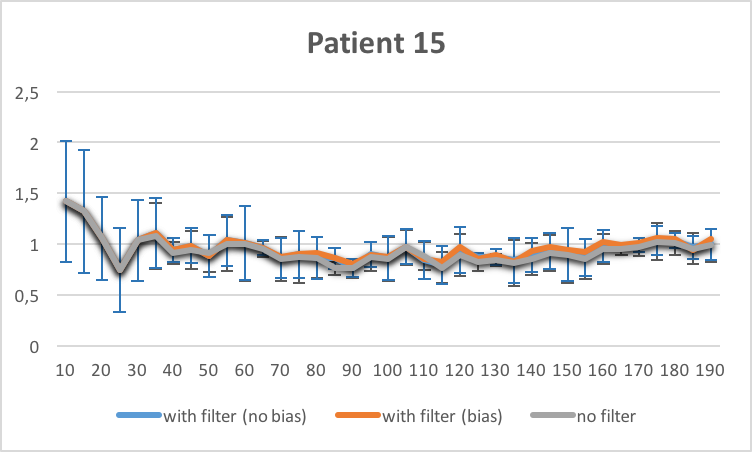}
}
\subfigure[Patient 16]{
\centering
\includegraphics[width=0.45\columnwidth]{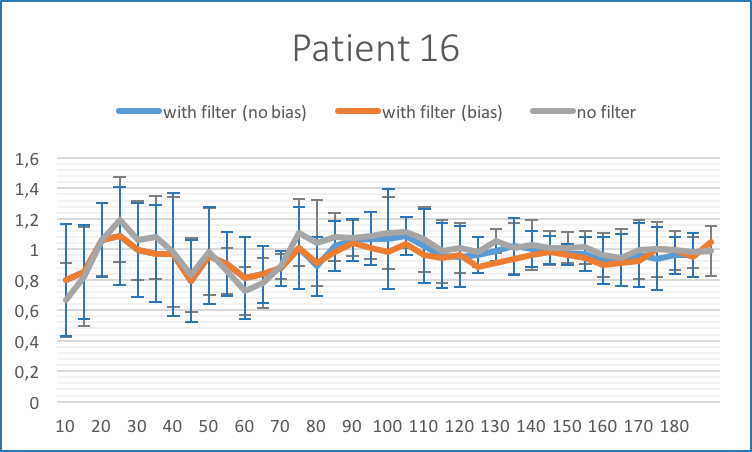}
}
\caption{5-fold cross validation with incremental training size.}
\label{fig:fcomb}
\end{figure}

\begin{figure}[ht]
\subfigure[Patient 8]{
\centering
\includegraphics[width=0.3\columnwidth]{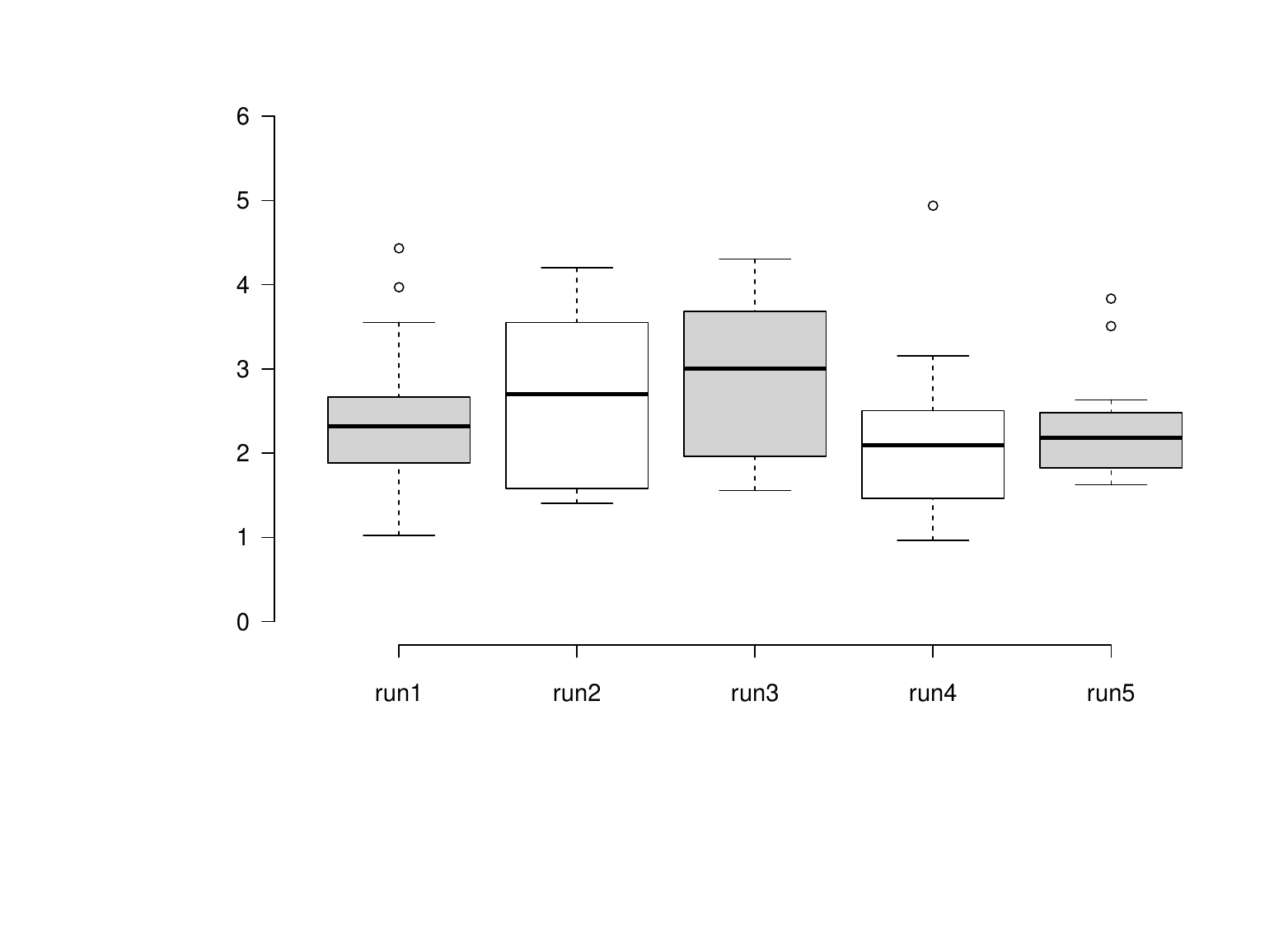}
}
\subfigure[Patient 10]{
\centering
\includegraphics[width=0.3\columnwidth]{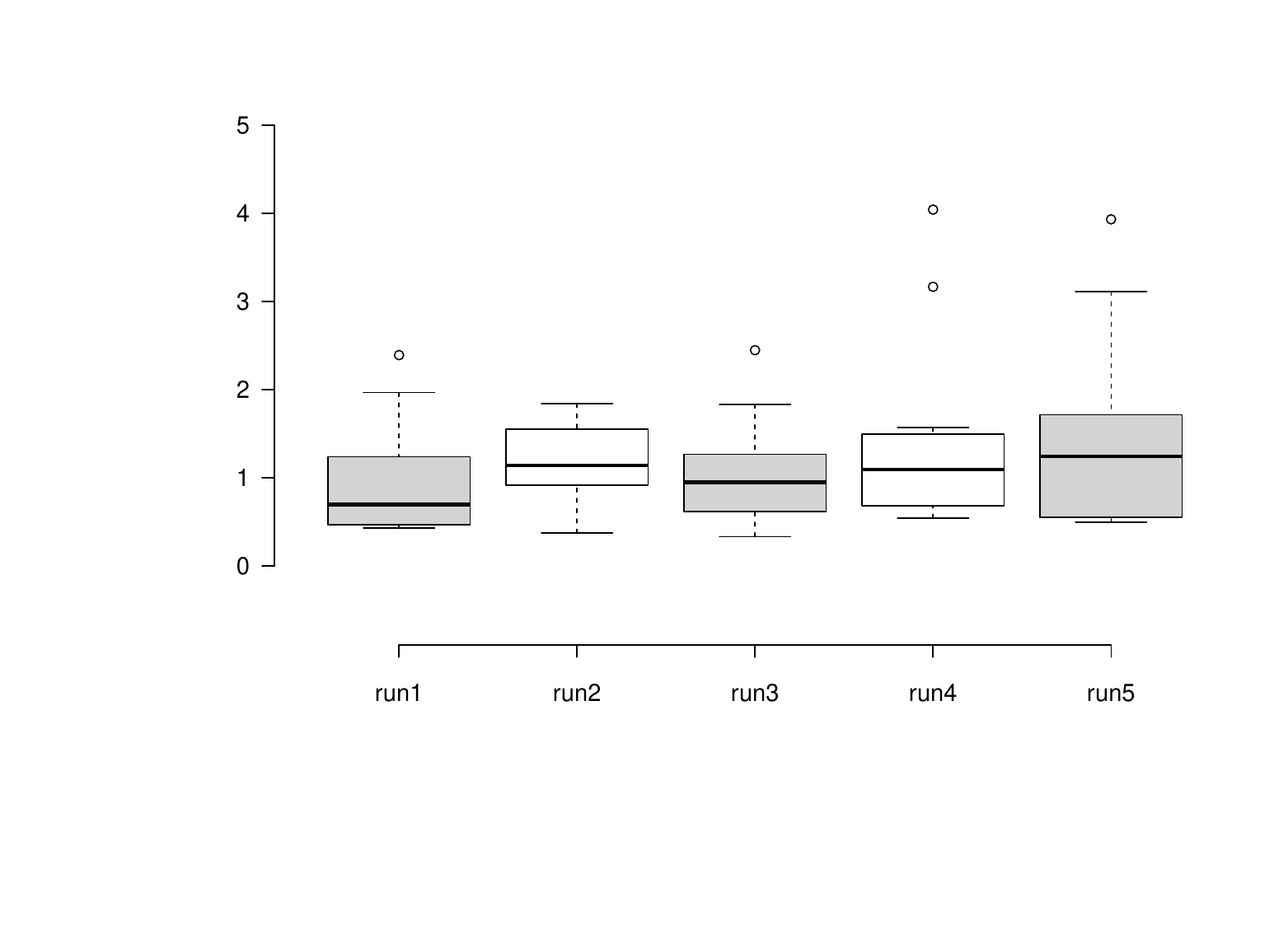}
}
\subfigure[Patient 11]{
\centering
\includegraphics[width=0.3\columnwidth]{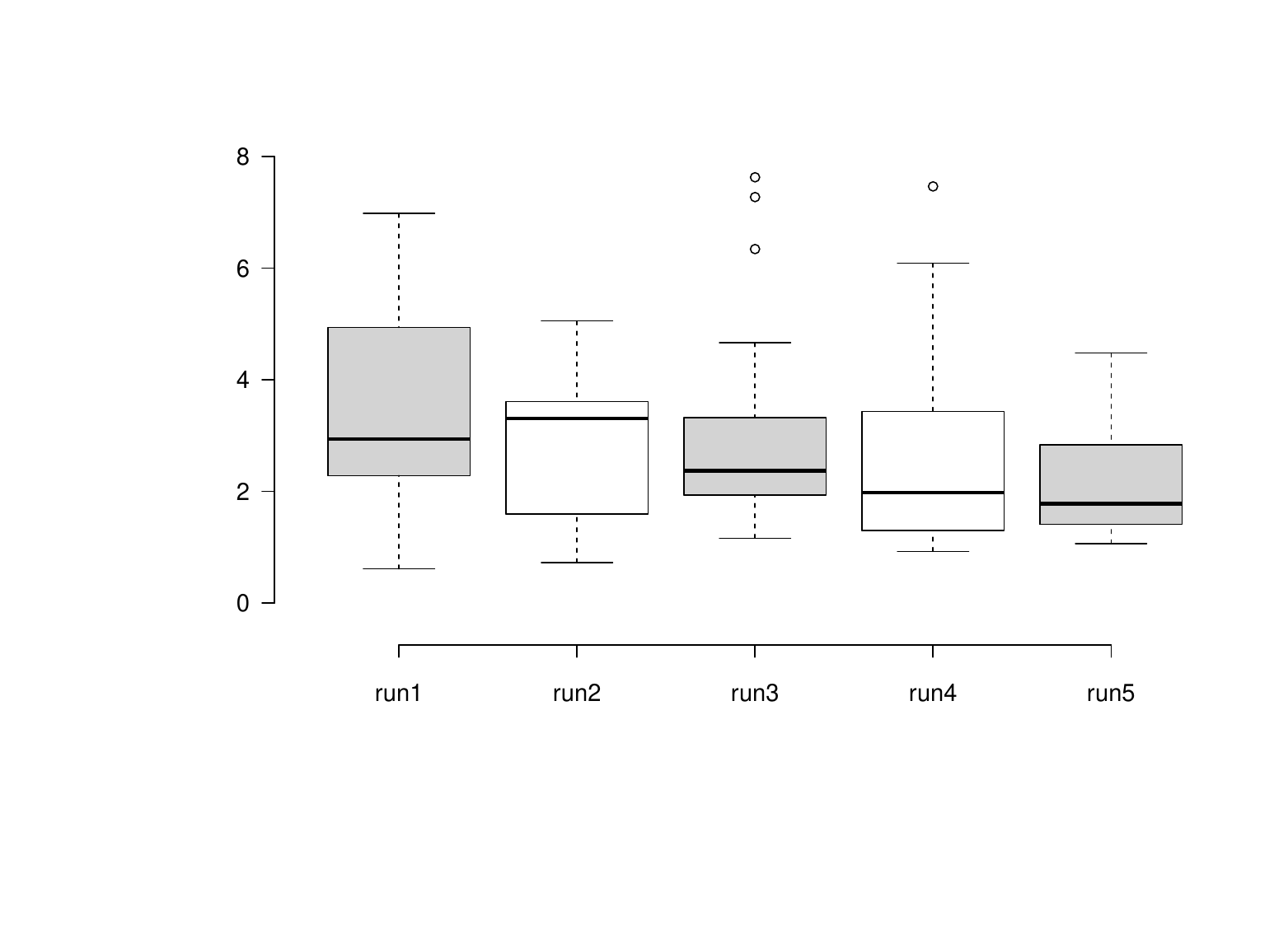}}\\
\subfigure[Patient 12]{
\centering
\includegraphics[width=0.3\columnwidth]{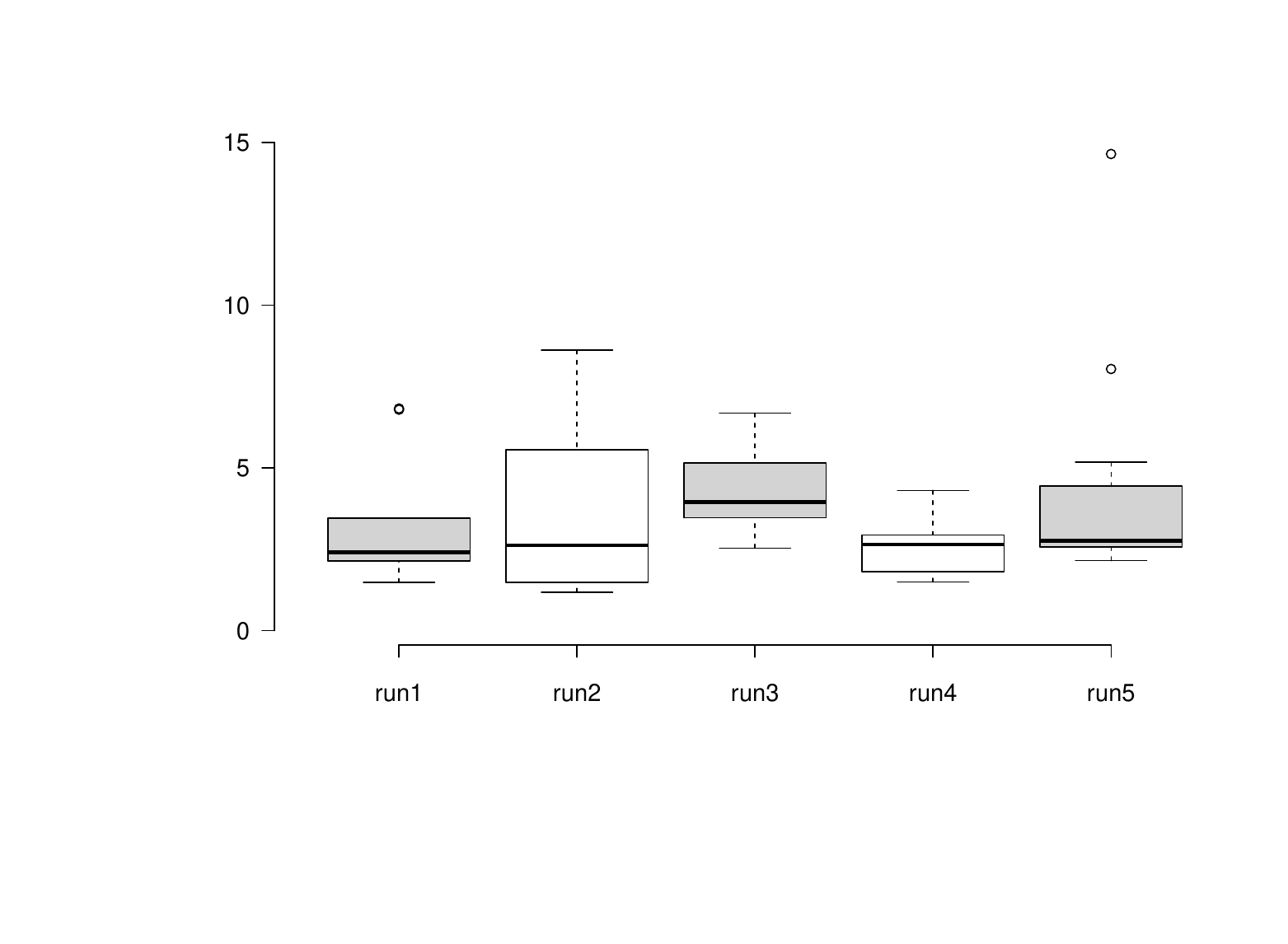}
}
\subfigure[Patient 13]{
\centering
\includegraphics[width=0.3\columnwidth]{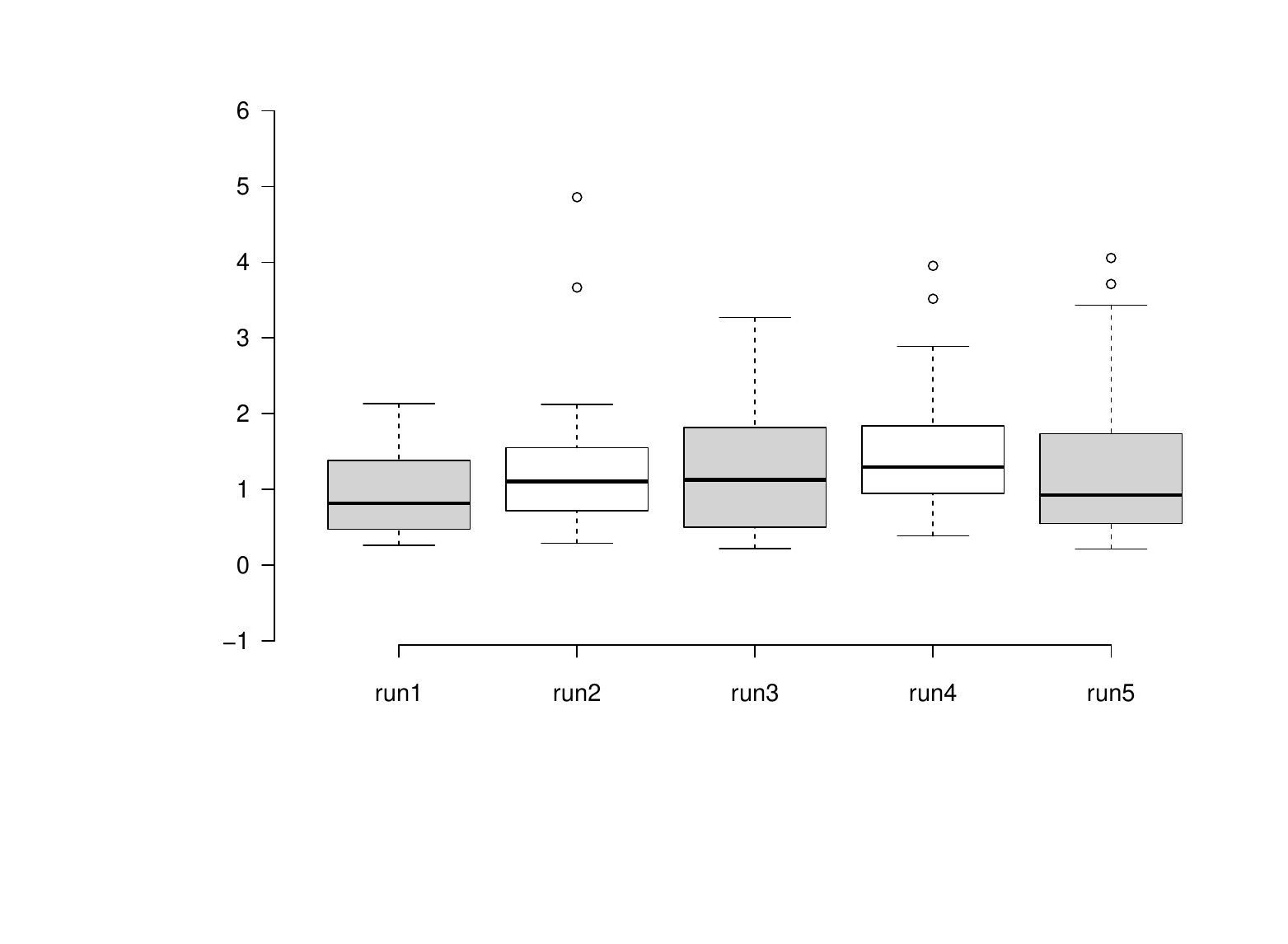}
}
\subfigure[Patient 14]{
\centering
\includegraphics[width=0.3\columnwidth]{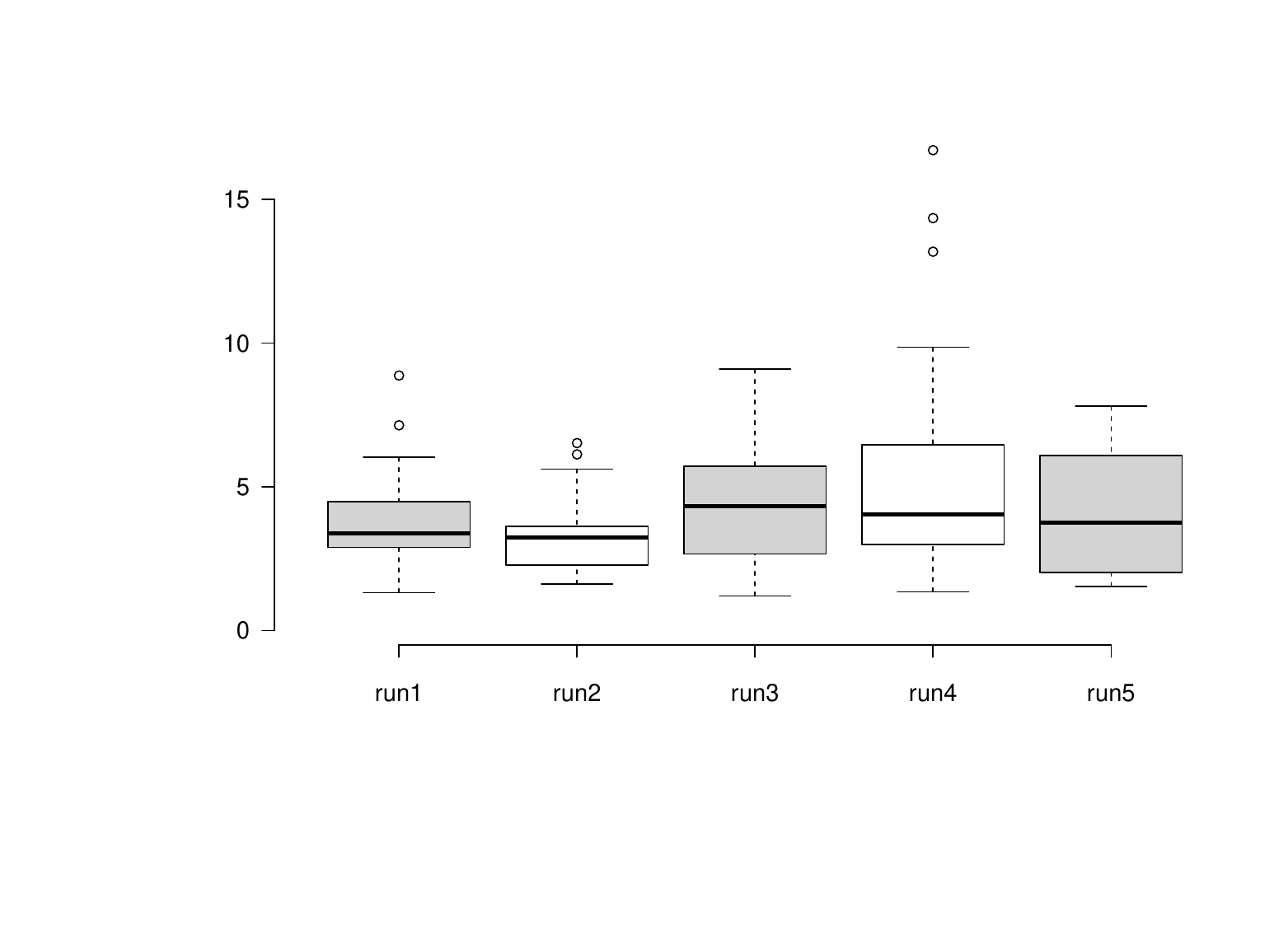}
}\\
\subfigure[Patient 15]{
\centering
\includegraphics[width=0.3\columnwidth]{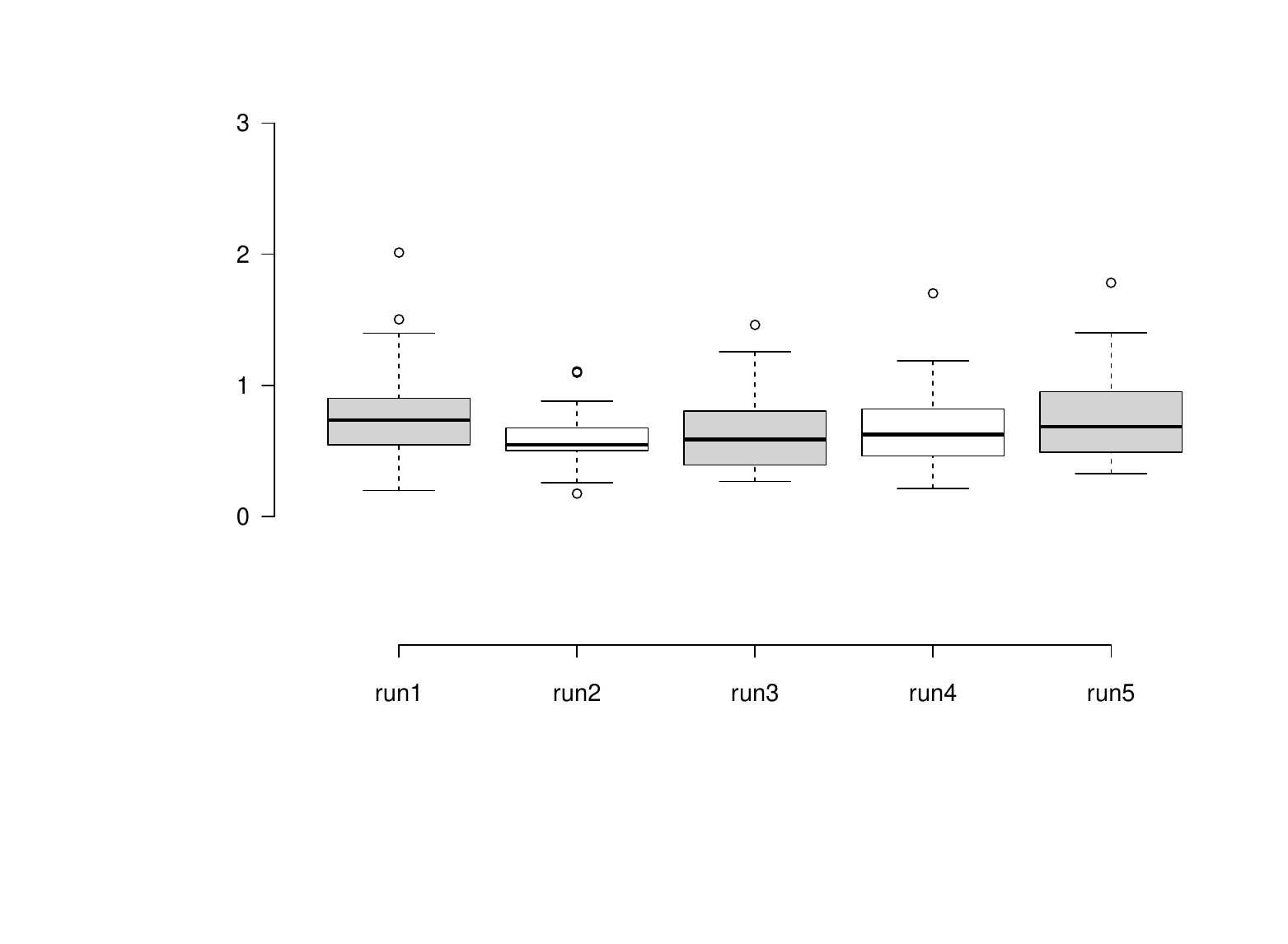}
}
\subfigure[Patient 16]{
\centering
\includegraphics[width=0.3\columnwidth]{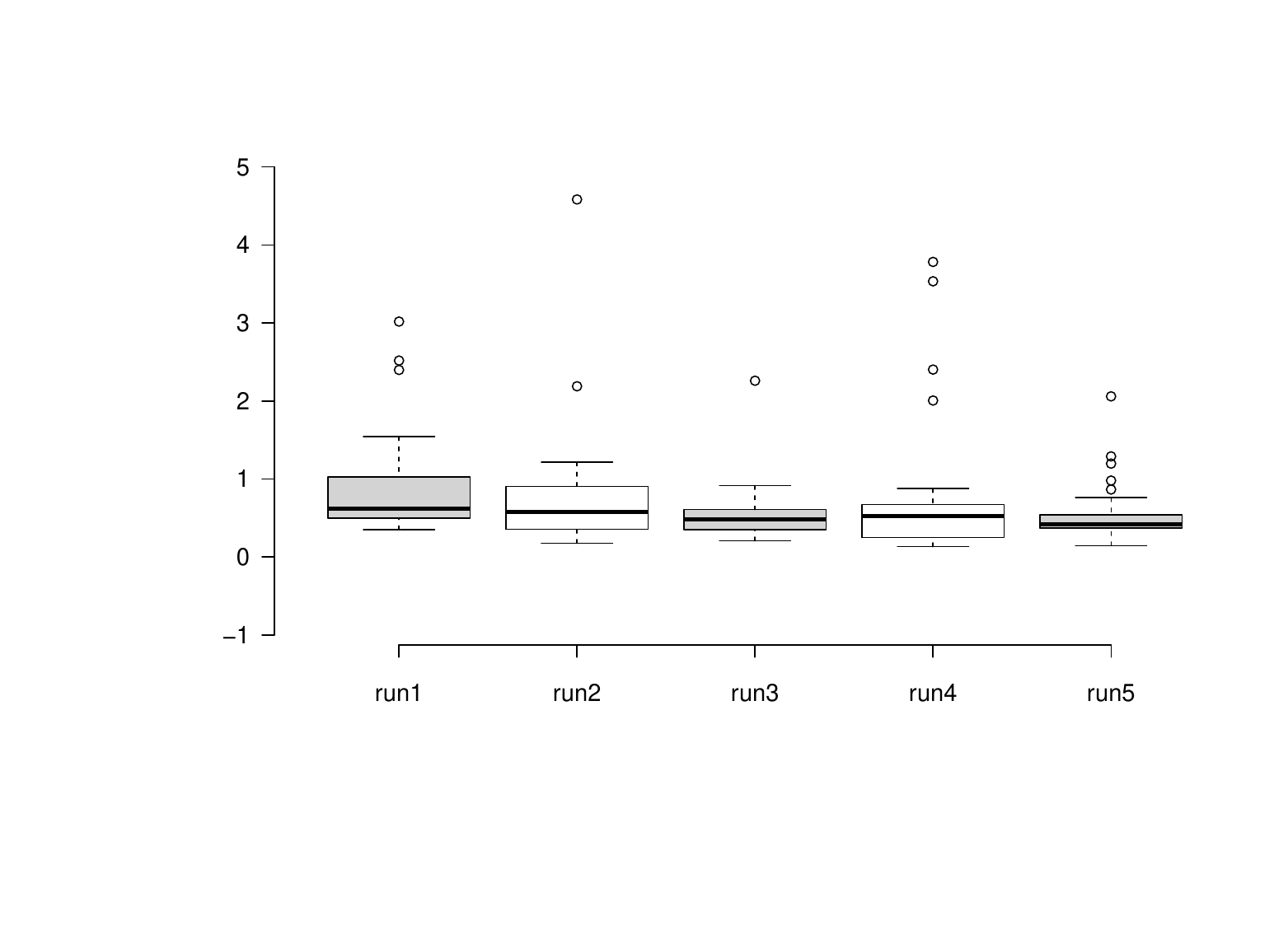}
}
\subfigure[Patient 17]{
\centering
\includegraphics[width=0.3\columnwidth]{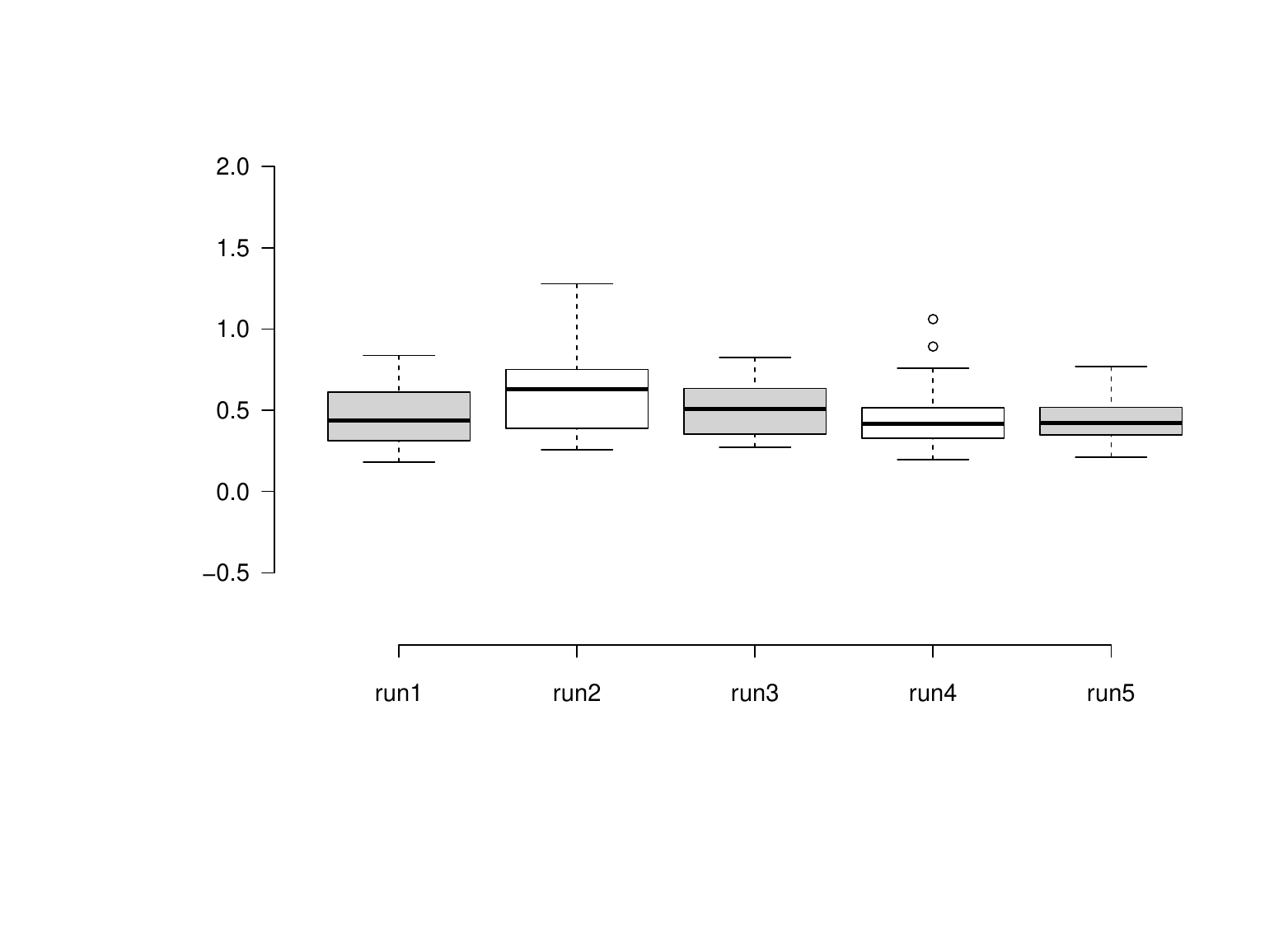}
}
\caption{Confidence distributions at each run of 5-fold CV for predicted BG using biased variance.}
\label{fig:disconfbias}
\end{figure}

\begin{figure}[ht]
\subfigure[Patient 8]{
\centering
\includegraphics[width=0.3\columnwidth]{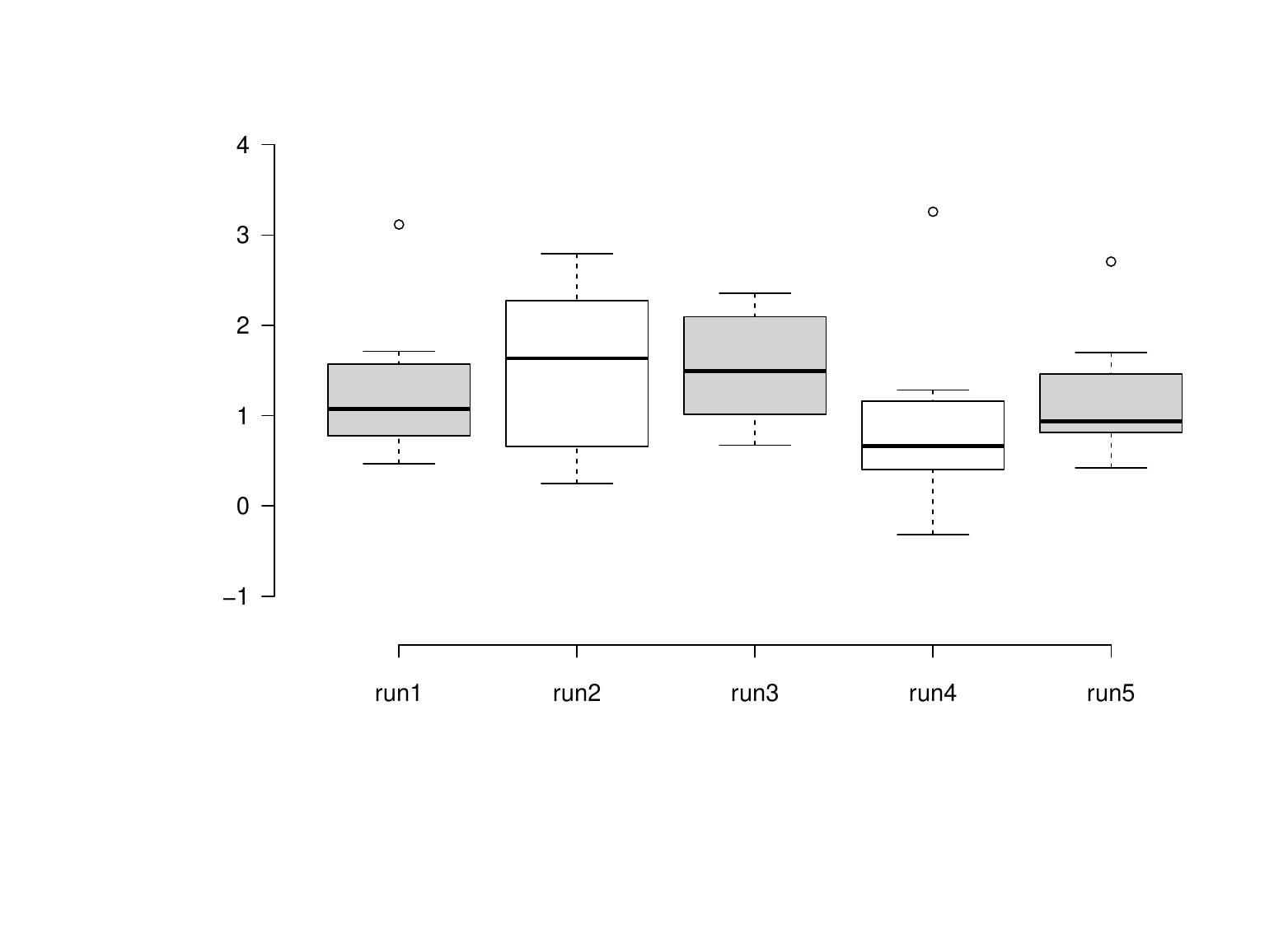}
}
\subfigure[Patient 10]{
\centering
\includegraphics[width=0.3\columnwidth]{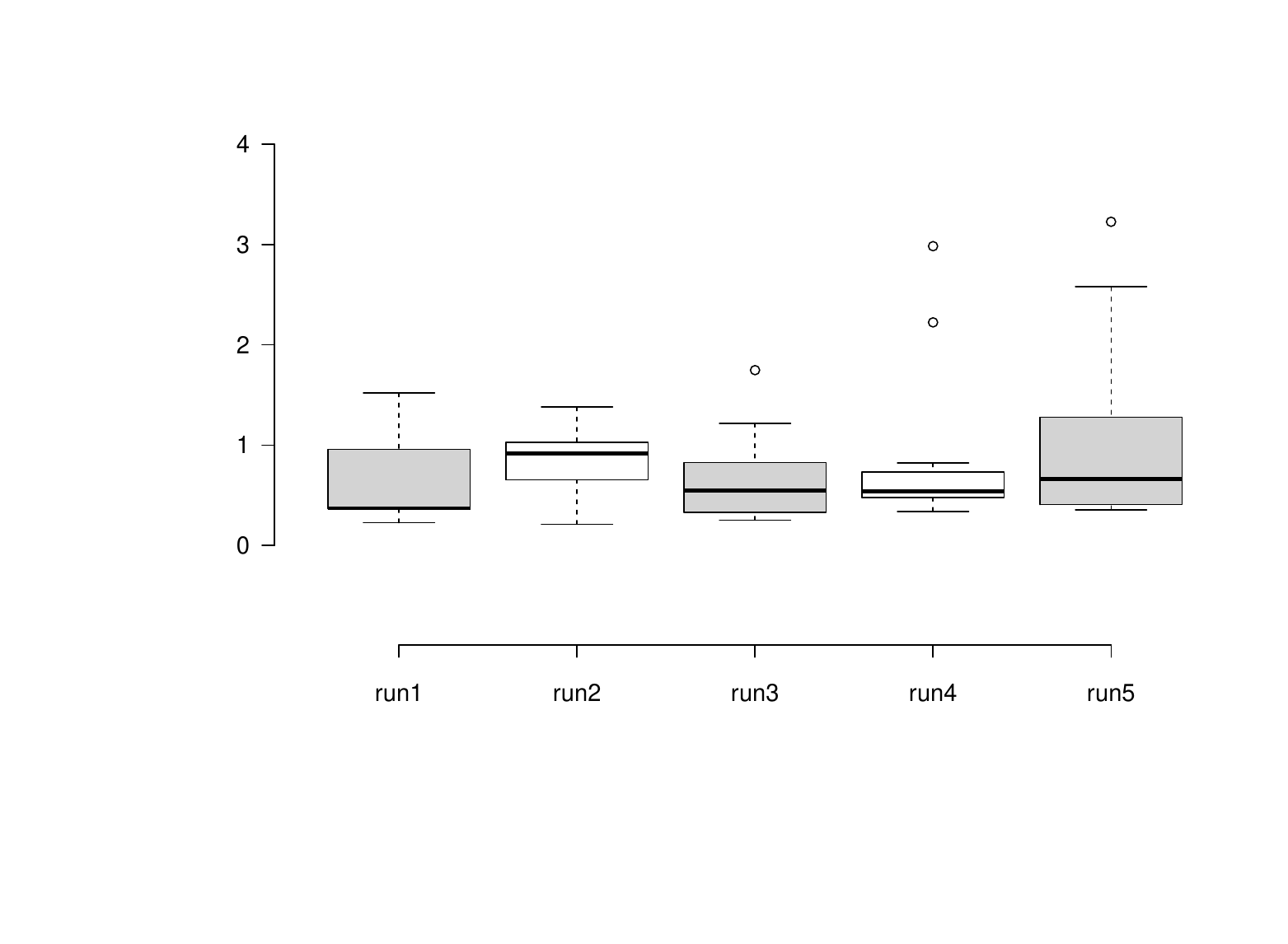}
}
\subfigure[Patient 11]{
\centering
\includegraphics[width=0.3\columnwidth]{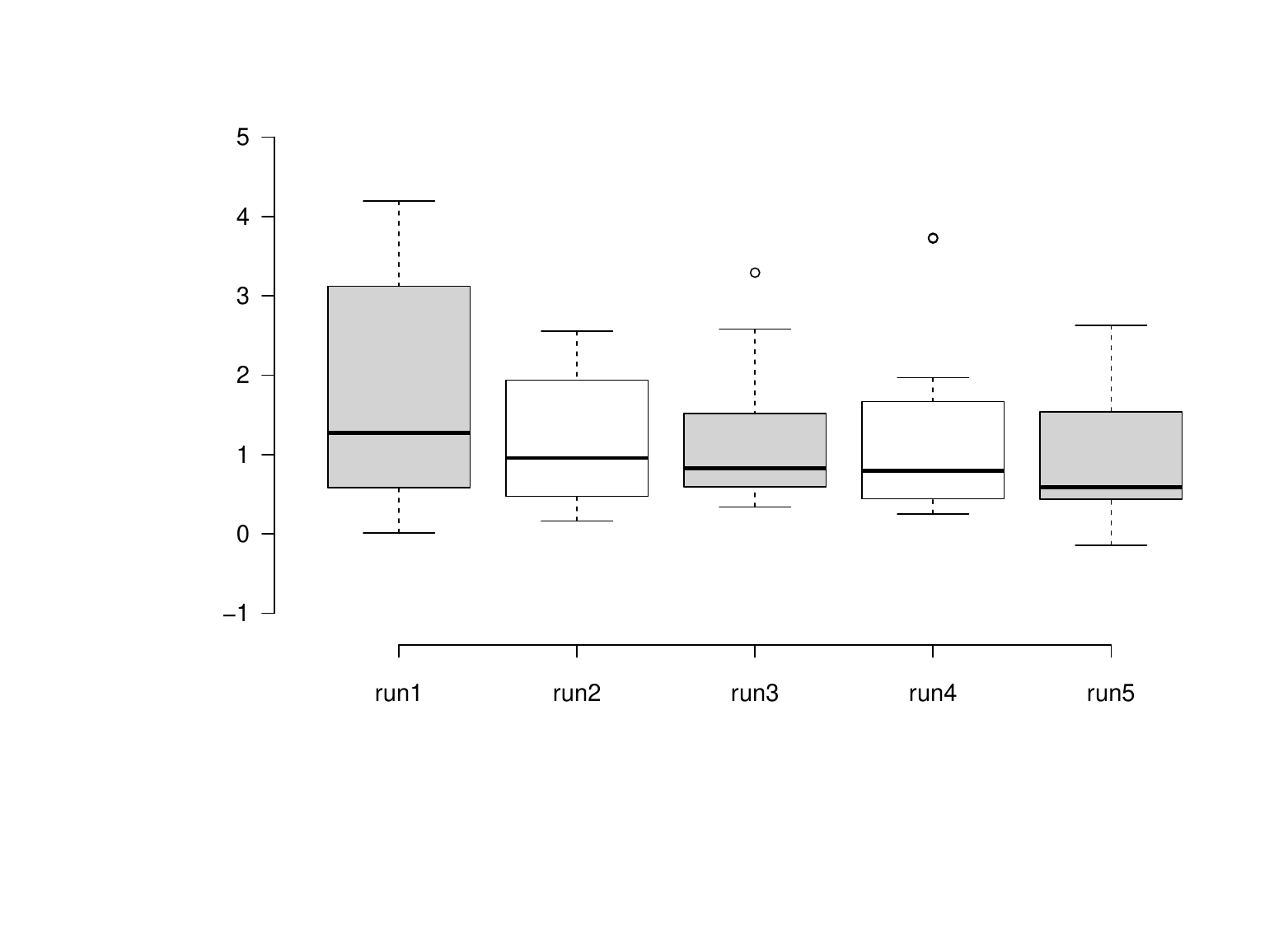}}\\
\subfigure[Patient 12]{
\centering
\includegraphics[width=0.3\columnwidth]{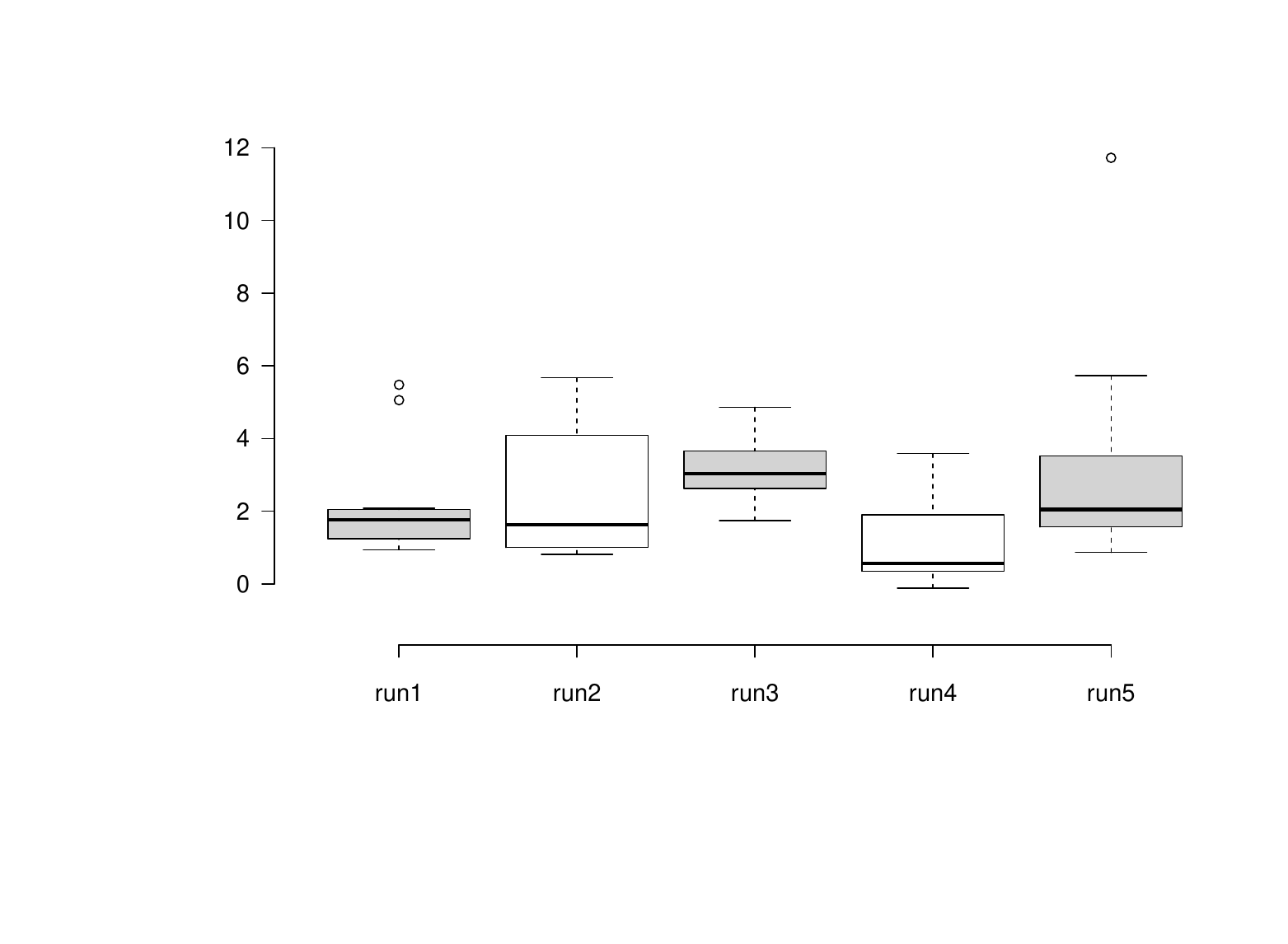}
}
\subfigure[Patient 13]{
\centering
\includegraphics[width=0.3\columnwidth]{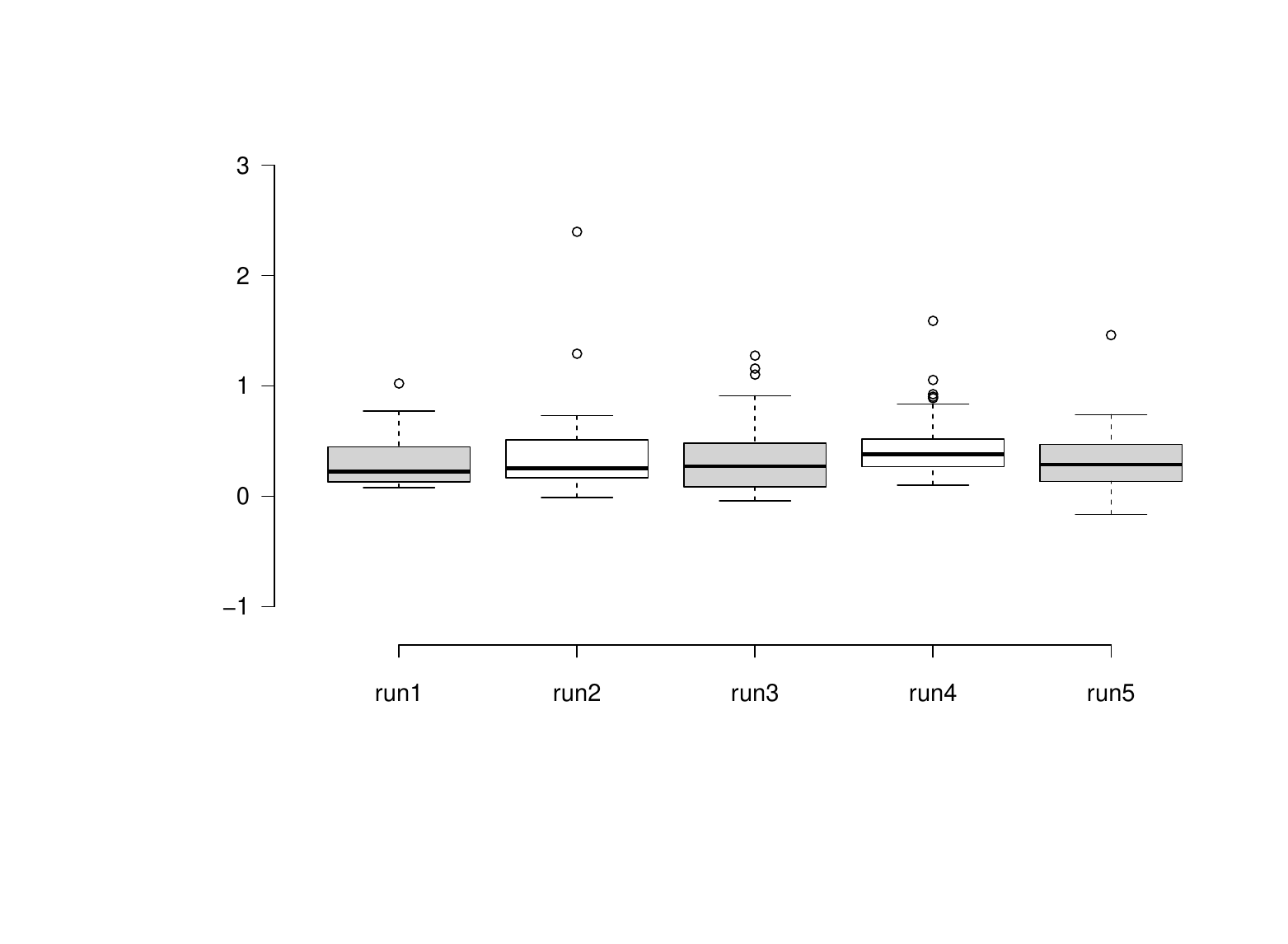}
}
\subfigure[Patient 14]{
\centering
\includegraphics[width=0.3\columnwidth]{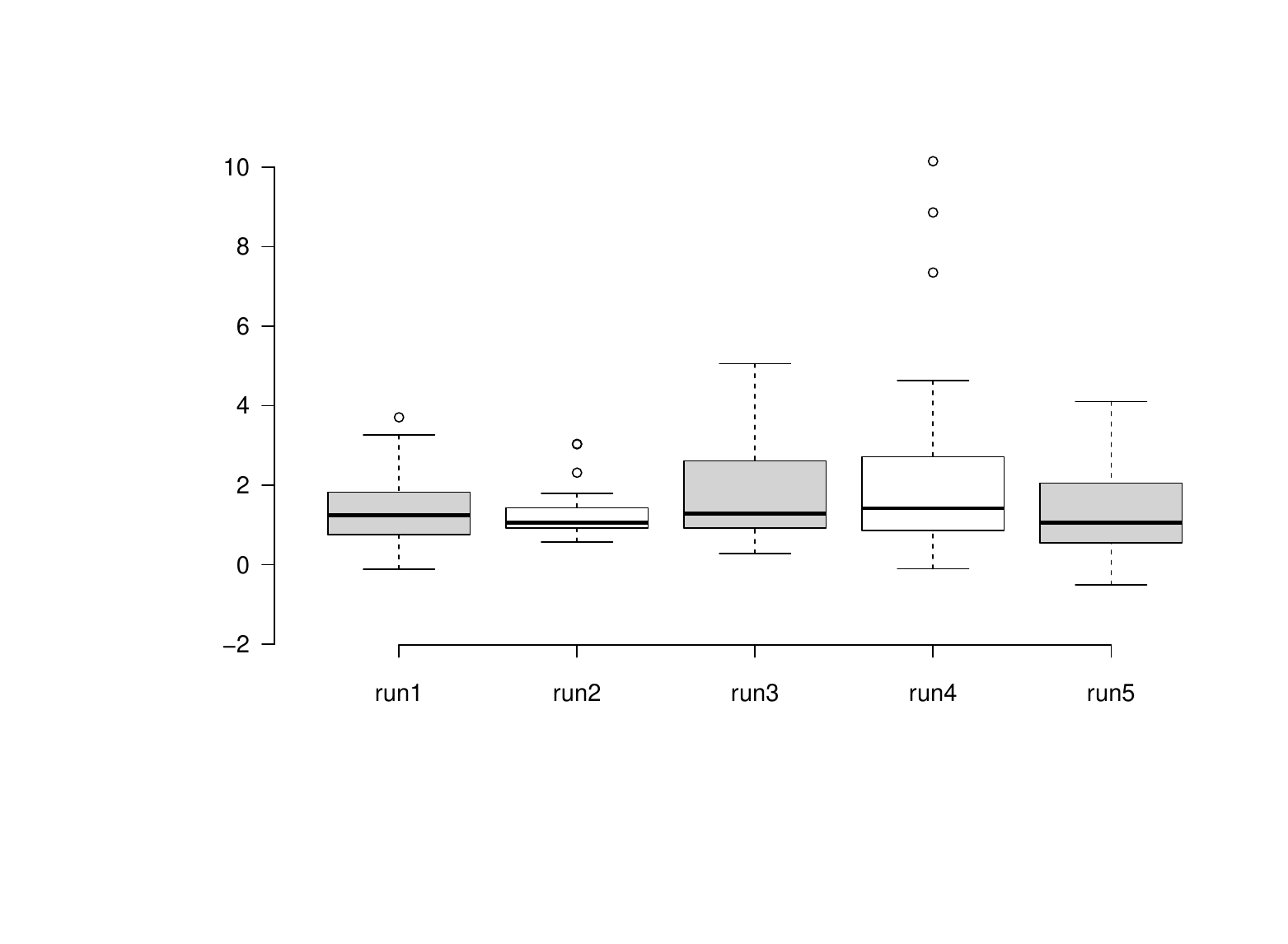}
}\\
\subfigure[Patient 15]{
\centering
\includegraphics[width=0.3\columnwidth]{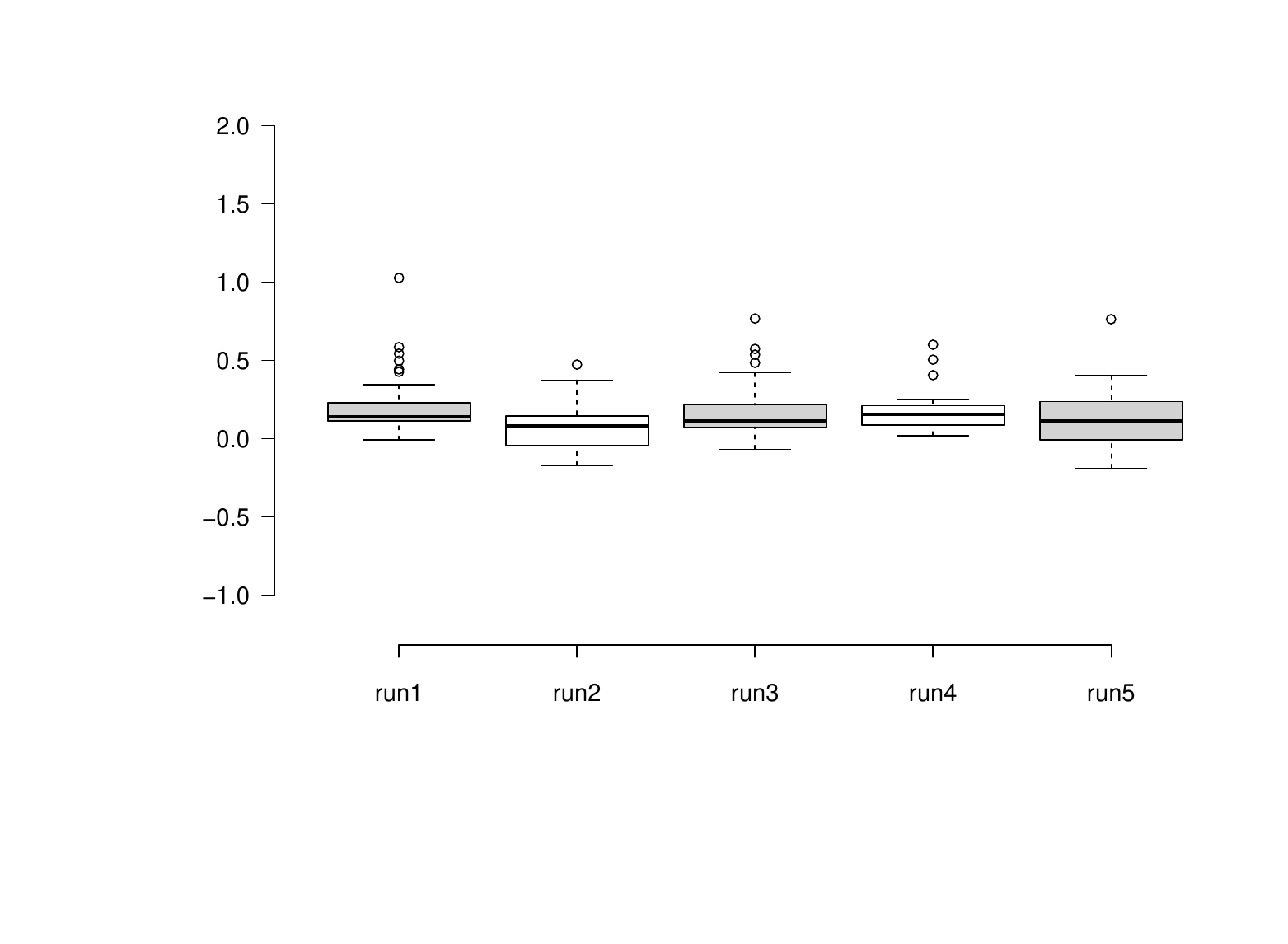}
}
\subfigure[Patient 16]{
\centering
\includegraphics[width=0.3\columnwidth]{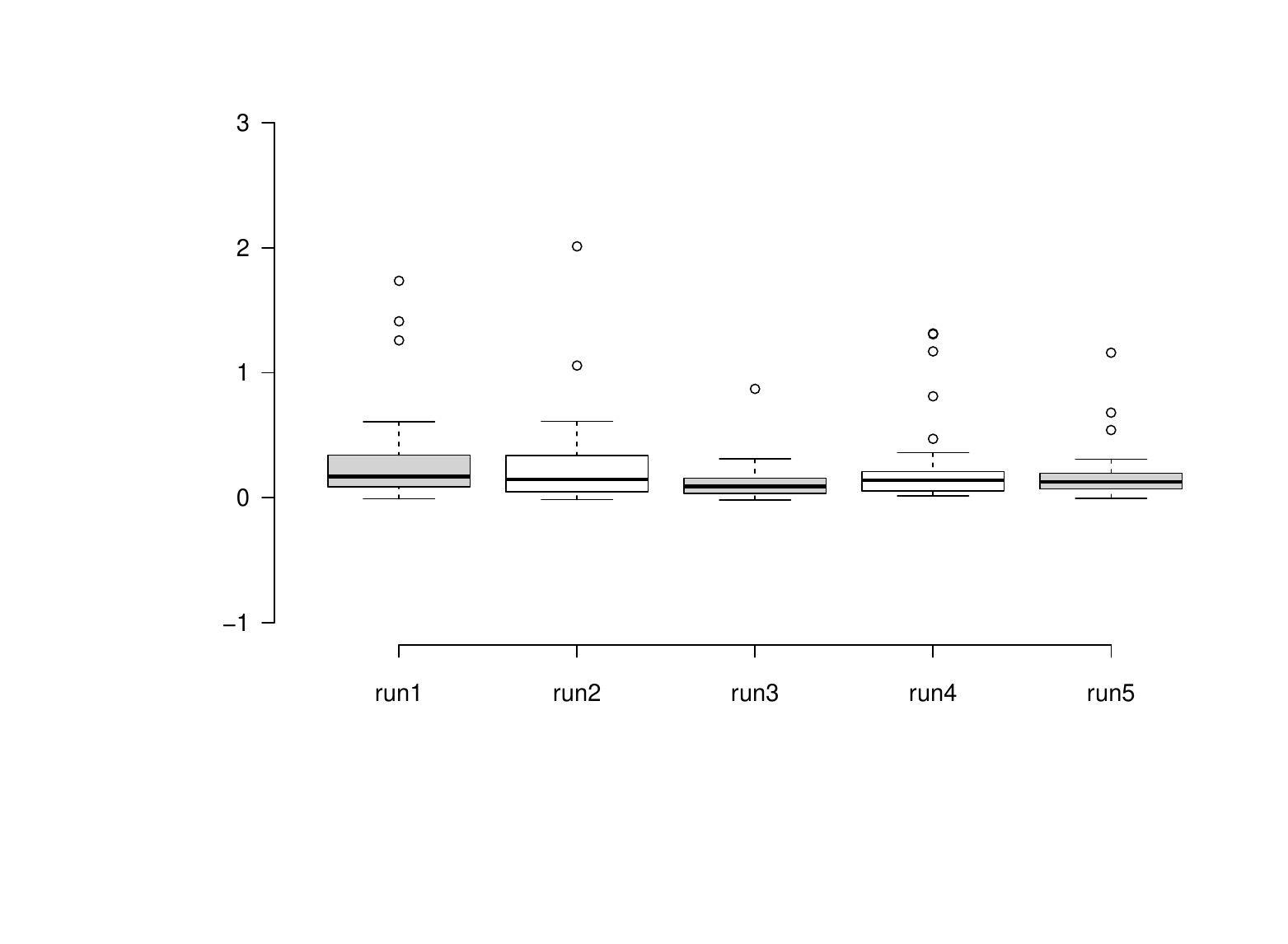}
}
\subfigure[Patient 17]{
\centering
\includegraphics[width=0.3\columnwidth]{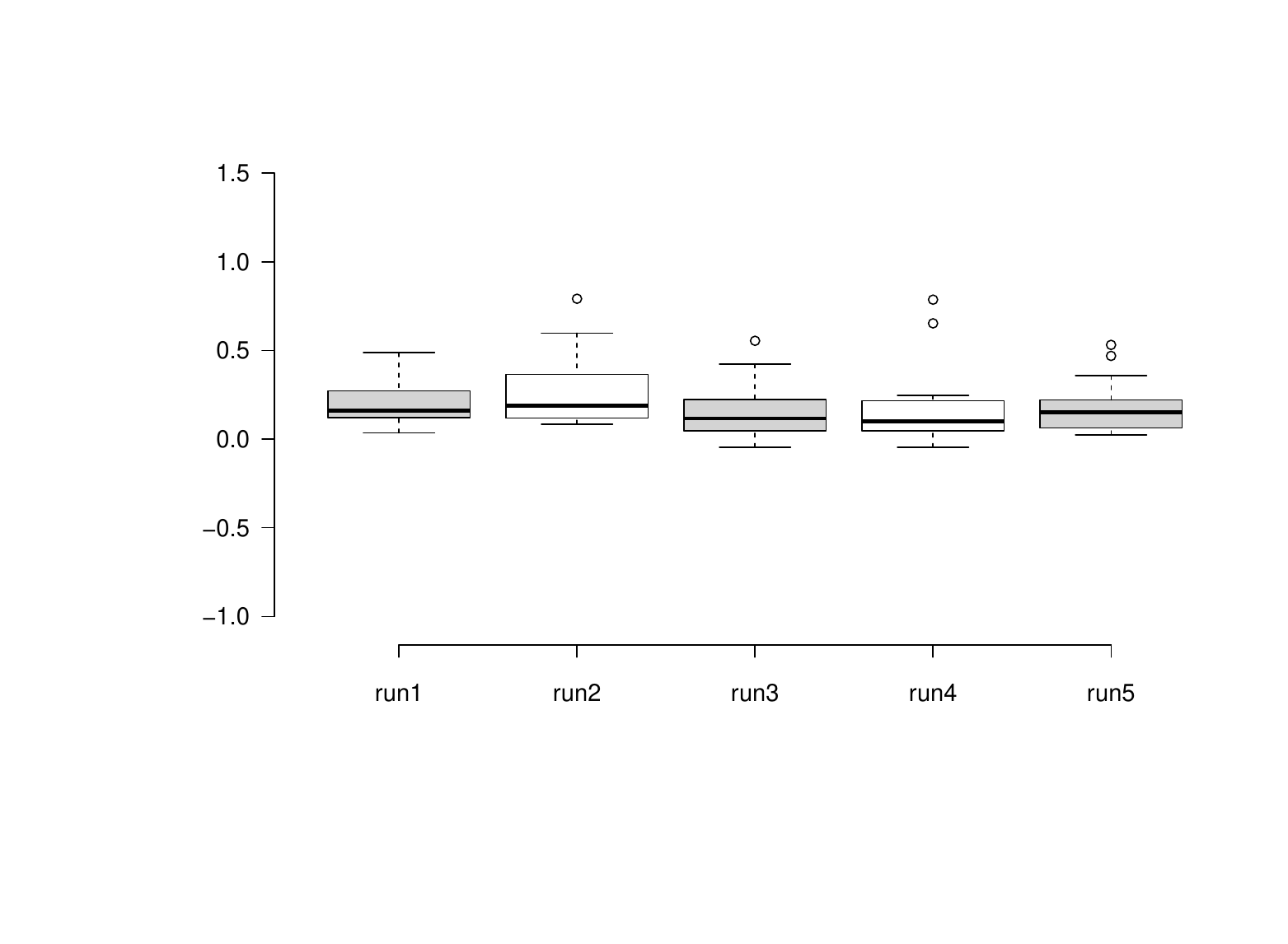}
}
\caption{Confidence distributions at each run of 5-fold CV for predicted BG using unbiased variance.}
\label{fig:disconf}
\end{figure}

\begin{figure}[H]
\subfigure[Patient 15]{
\centering
\includegraphics[width=0.45\columnwidth]{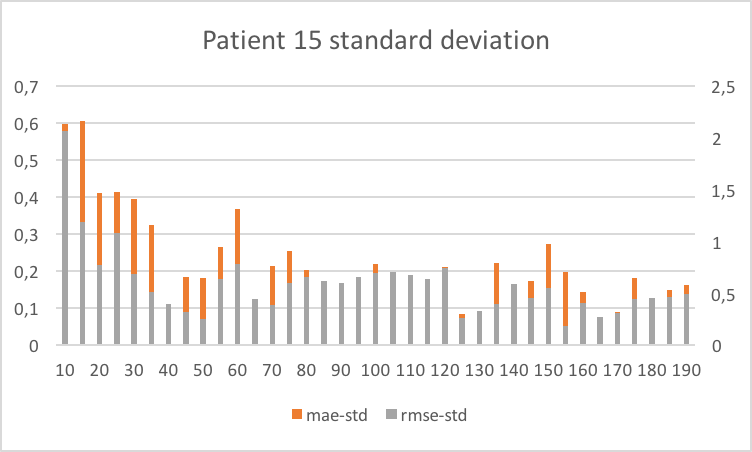}
}
\subfigure[Patient 17]{
\centering
\includegraphics[width=0.45\columnwidth]{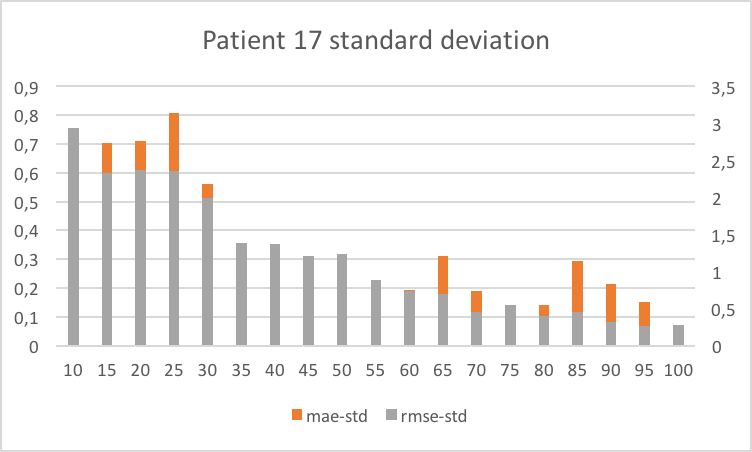}
}
\caption{Standard deviation with incremental training size.}
\label{fig:std}
\end{figure}

\begin{table}[ht]
\centering
\caption{Average performance of different filtering approaches for all patients.}
\label{tab:filtersum}
\small{
\begin{tabular}{|l|l|l|l|l|l|}
\hline
\textbf{Model}                                 & \textbf{\# predictions} & \textbf{MAE}  & \textbf{MdAE} & \textbf{RMSE}  & \textbf{SMAPE}  \\  \hline
\textbf{rf}                                    & 42 & 2.58       & 2.27          & 12.05         & 29.98        \\ \hline
\textbf{et}                                    & 42 & 2.55       & 2.16        & 12.15        & 29.56       \\ \hline
\textbf{rf + sanity filter}                    & 16 & 2.22      & 2.01               & 8.80        & 28.10                 \\ \hline
\textbf{et + sanity filter}                    & 16 & 2.29       & 2.06         & 9.01         & 29.36          \\ \hline
\textbf{rf + sanity + stability filter} & 15 & 2.22 & 1.92 & 8.71        & 27.82          \\ \hline
\textbf{rf + stability filter}                 & 24 & \textbf{1.92} & \textbf{1.77}                & \textbf{7.57} & \textbf{22.65} \\ \hline
\end{tabular}
}
\end{table}

\section{Conclusion}
We studied the predictability of machine-learning models in the scenarios of non-continuous blood glucose tracking. Additionally, we studied the stability and robustness of the learned model over time. We show that Random Forest and Extra Tree ensemble-based models are the most suitable models for this case, as they can account for the outliers as well as overfitting problems when the data are limited. Our further study on the prediction confidence show that the model can give reliable predictions after acquiring 25-30 instances.

\noindent\textbf{Acknowledgements.} {This work was partially funded by the German Federal Ministry of Education and Research (BMBF) under project GlycoRec (16SV7172).}
\bibliographystyle{abbrv}

\bibliography{paperb}

\begin{figure}[h]
\subfigure[Patient 8 - bias]{
\centering
\includegraphics[width=0.45\columnwidth]{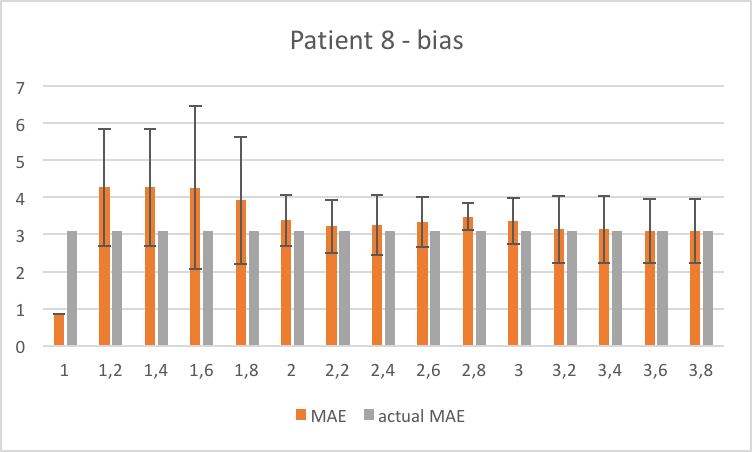}
}
\subfigure[Patient 8 - no bias]{
\centering
\includegraphics[width=0.45\columnwidth]{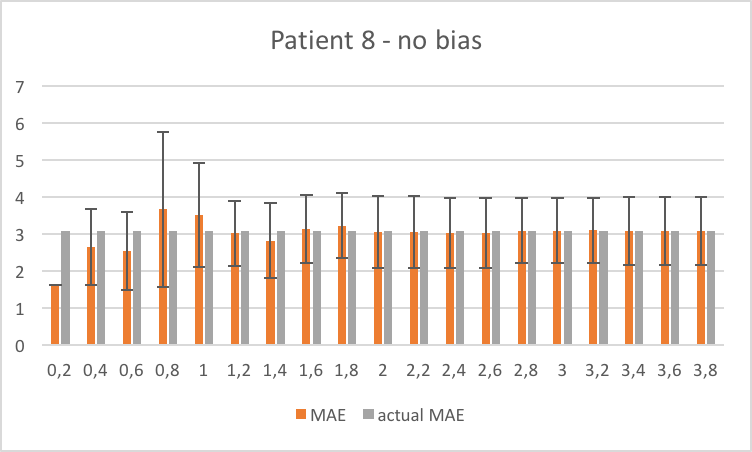}
}\\
\subfigure[Patient 10 - bias ]{
\centering
\includegraphics[width=0.45\columnwidth]{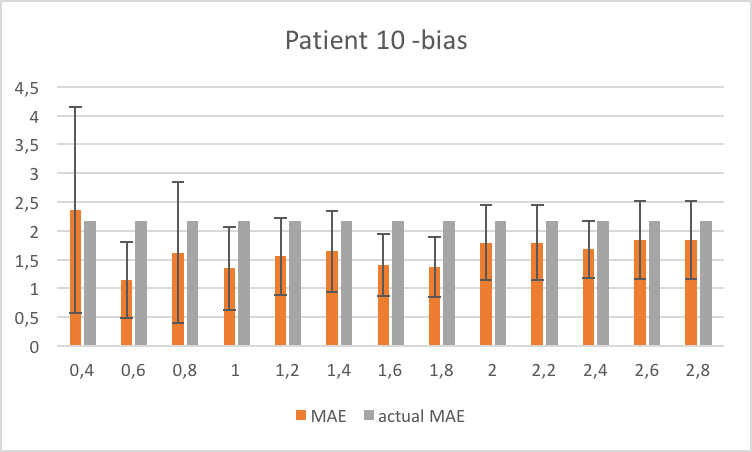}}
\subfigure[Patient 10 - no bias]{
\centering
\includegraphics[width=0.45\columnwidth]{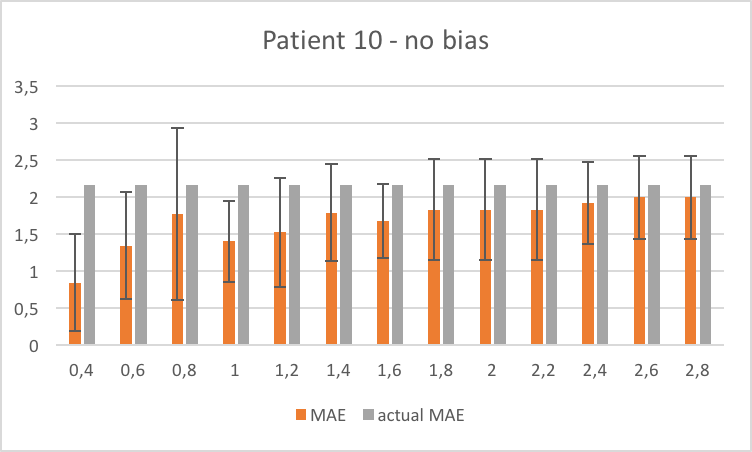}
}\\
\subfigure[Patient 13 - bias]{
\centering
\includegraphics[width=0.45\columnwidth]{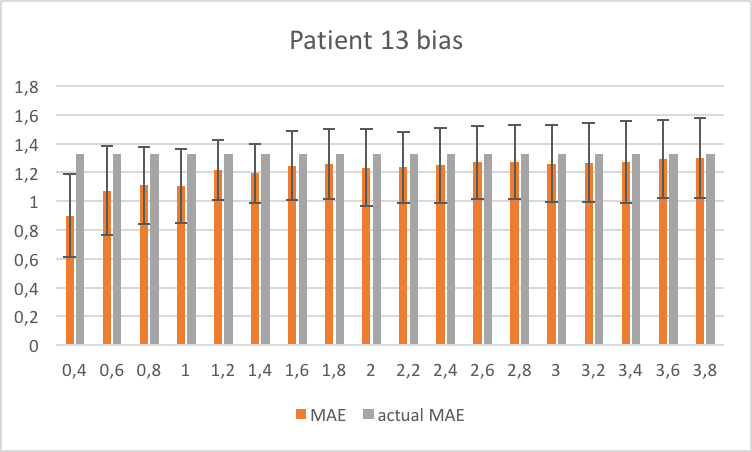}
}
\subfigure[Patient 13 - no bias]{
\centering
\includegraphics[width=0.45\columnwidth]{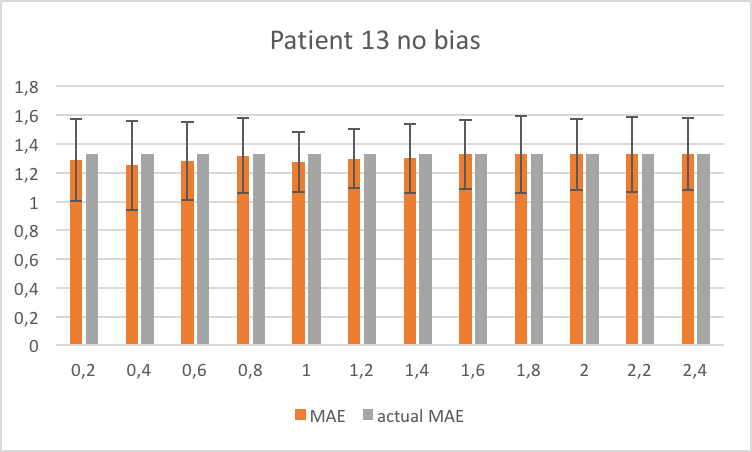}
}\\
\subfigure[Patient 14 - bias]{
\centering
\includegraphics[width=0.45\columnwidth]{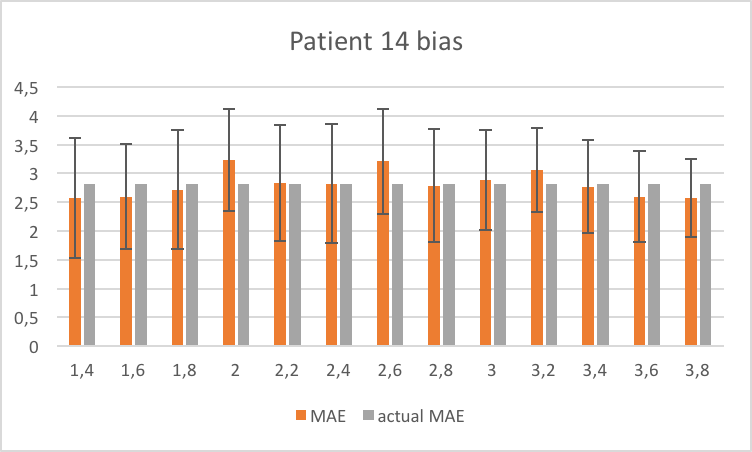}
}
\subfigure[Patient 14 - no bias]{
\centering
\includegraphics[width=0.45\columnwidth]{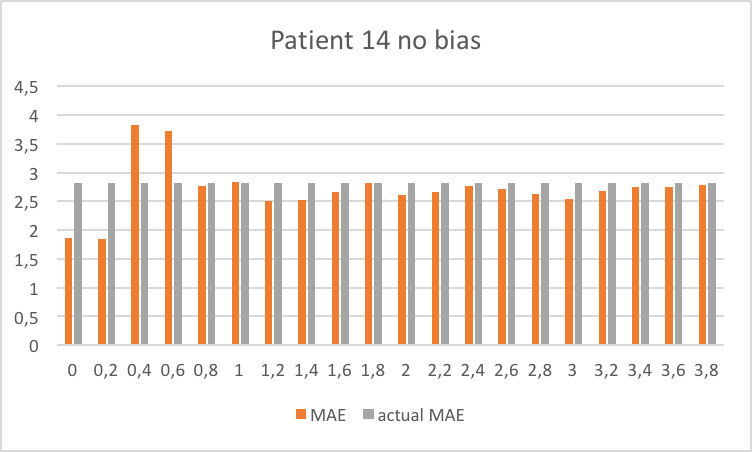}
}
\caption{Performance with confidence filtering threshold (x-axis) for some patients, MAE is when filtering is applied.}
\label{fig:cthresh}
\end{figure}

\begin{figure}[h]
\subfigure[Patient 8]{
\centering
\includegraphics[width=0.45\columnwidth]{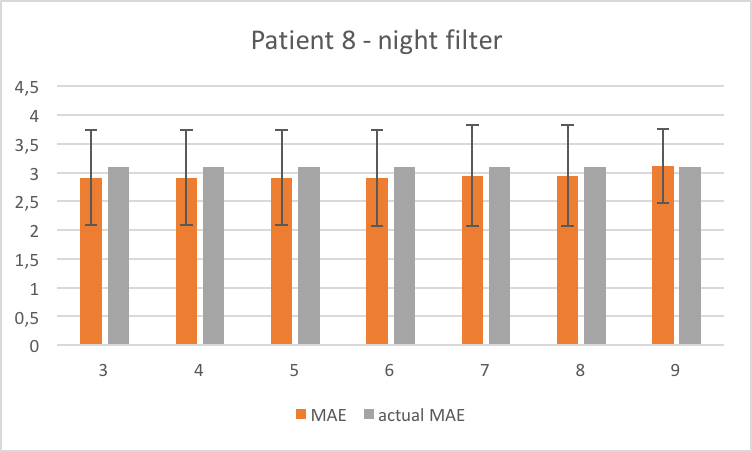}
}
\subfigure[Patient 13]{
\centering
\includegraphics[width=0.45\columnwidth]{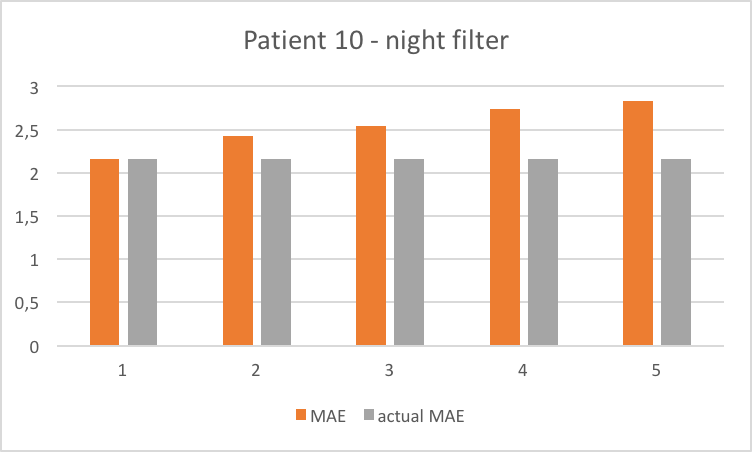}
}\\
\subfigure[Patient 15]{
\centering
\includegraphics[width=0.45\columnwidth]{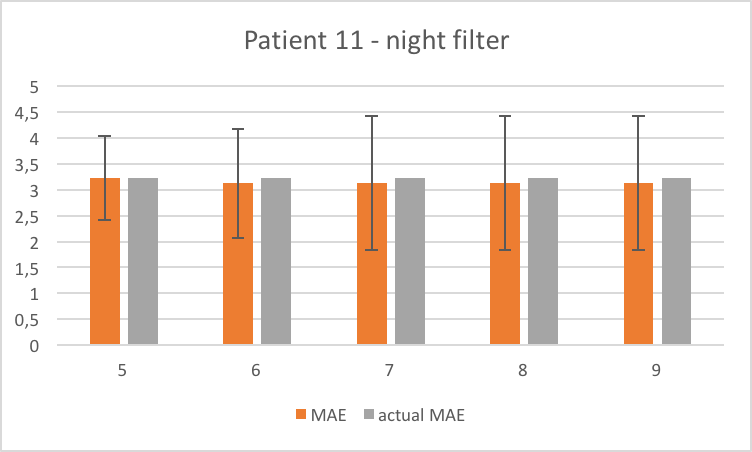}
}
\subfigure[Patient 16]{
\centering
\includegraphics[width=0.45\columnwidth]{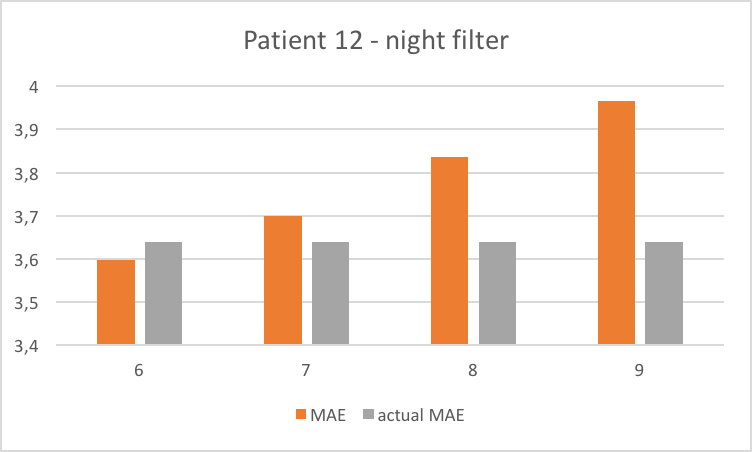}
}
\caption{Performance with night time filtering, x-axis is the night time, MAE is when filtering is applied.}
\label{fig:night}
\end{figure}

\end{document}